\documentclass[aps, twocolumn, superscriptaddress, showpacs, nofootinbib, longbibliography]{revtex4-1}

\usepackage[utf8]{inputenc}
\usepackage[T1]{fontenc}
\usepackage{ae,aecompl} 
\usepackage{graphicx}
\usepackage{amsmath}
\usepackage{color}
\usepackage{amssymb}
\usepackage{latexsym}
\usepackage{wasysym}
\usepackage{psfrag}
\usepackage{ifthen}
\usepackage[citecolor=blue,colorlinks=true]{hyperref}
\usepackage{float}
\usepackage[utf8]{inputenc}
\usepackage{lineno}

\usepackage{aas_macros}
\usepackage{xcolor}
\usepackage{booktabs}
\usepackage[normalem]{ulem}
\usepackage{placeins}
\usepackage{subcaption}
\usepackage{cleveref}

\definecolor{bbsalmon}{rgb}{1.0, 0.47, 0.42}
\definecolor{datablue}{rgb}{0.0, 0.0, 1.0}

\newcommand{\dl}[1]{\textcolor{black}{#1}}

\begin{document}

\preprint{APS/123-QED}

\title{Utilizing Gaussian mixture models in all-sky searches for short-duration gravitational wave bursts}
%\title{Gaussian mixture modeling utilization in all-sky search for short-duration gravitational wave burst} %Application of Gaussian mixture modeling in all-sky short-duration gravitational-wave burst search
% during the Advanced LIGO first half of the third observing run }

\author{Dixeena Lopez}

\affiliation{Physik-Institut, University of Zurich, Winterthurerstrasse 190, 8057 Zurich, Switzerland }
\author{V. Gayathri}
\affiliation{Department of Physics, University of Florida, PO Box 118440, Gainesville, FL 32611-8440, USA}
\author{Archana Pai}
\affiliation{Department of Physics, Indian Institute of Technology Bombay, Powai, Mumbai 400 076, India}
\author{Ik Siong Heng}
\affiliation{SUPA, School of Physics and Astronomy, University of Glasgow, Glasgow G12 8QQ, United Kingdom}
\author{Chris Messenger}
\affiliation{SUPA, School of Physics and Astronomy, University of Glasgow, Glasgow G12 8QQ, United Kingdom}
\author{Sagar Kumar Gupta}
\affiliation{Department of Physics, Indian Institute of Technology Bombay, Powai, Mumbai 400 076, India}

\begin{abstract}
Coherent WaveBurst is a generic, multidetector gravitational wave burst search based on the excess power approach.  The coherent WaveBurst algorithm currently employed in the all-sky short-duration gravitational wave burst search uses a conditional approach on selected attributes in the multidimensional event attribute space to distinguish between noisy events from that of astrophysical origin. We have been developing
a supervised machine learning approach based on the Gaussian mixture modeling to model the
attribute space for signals as well as noise events to enhance the probability of burst detection~\cite{gmm}. We
further extend the GMM approach to the all-sky short-duration coherent WaveBurst search as a postprocessing step
on events from the first half of the third observing run (O3a). We show an improvement in sensitivity
to generic gravitational wave burst signal morphologies as well as the astrophysical source such as
core-collapse supernova models due to the application of our Gaussian mixture model approach to
coherent WaveBurst triggers. The Gaussian mixture model method recovers the gravitational wave
signals from massive compact binary coalescences identified by coherent WaveBurst targeted for binary black holes in GWTC-2,
with better significance than the all-sky coherent WaveBurst search. No additional significant gravitational wave
bursts are observed.

\end{abstract}

\maketitle

\section{INTRODUCTION}
%comment: Brief  update about 3rd observing run
\dl{The Gravitational wave (GW) catalogs of signals from the compact binary coalescence (CBC), present in the first half of the third observing run (O3a), had been recently released \cite{gwtc2,gwtc2_1} \footnote{The third observing run (O3) of the Advanced LIGO \cite{LIGO-det} and Advanced Virgo \cite{Virgo-det} (April 1, 2019, to March 27, 2020) detectors was divided into two segments as O3a run  (April 1, 2019 - October 1, 2019,)  and O3b (November 1, 2019 - March 27, 2020). } This includes intermediate black hole (IMBH) system GW190521 \cite{GW190521}, the merger of two compact objects with the unequal mass ratio GW190814 \cite{GW190814}, the inspiral of a binary neutron star (BNS) system, GW190425 \cite{GW190425}. Further,  the first confident observation of GW signal from a neutron star-black hole (NSBH) binary coalescence was made during O3b run \cite{NSBH,gwtc3}.}

Besides CBC signals, the current ground-based detectors are also sensitive to detecting GW short-duration bursts, such as signals from the core-collapse supernova (CCSN), cosmic strings, non-linear memory effects and isolated Neutron stars \cite{ccsn_ligo,cosmic_string,memory,magnetar,ns_glitch}. The LIGO-Virgo-Kagra collaboration has been actively searching for GW bursts signals since the fifth science run.  The all-sky burst search for short-duration transient  includes the burst signal duration of up to a few seconds \cite{allskyfith,allskysixth,allakyo1,alskyo2,allskyo3}. This search is identified as the unmodeled  search, sensitive to detect GW signals of different morphologies, including GW signals from the CBC system.

%comment:  Brief  update about  cWB  PP APPROACH
\dl{The morphology independent coherent WaveBurst (cWB) short-duration algorithm identifies triggers with short-duration excess power in the time-frequency domain \cite{cwb_2008}. The cWB made a significant contribution in the detection of the first GW signal from the binary black hole (BBH) merger GW150914  \cite{gw150914_cwb}  and the intermediate-mass black hole (IMBH) merger event \cite{gw190425_cwb}.  In cWB,  each trigger is associated with multiple attributes. The cWB  adopts an empirical classification of noise-based glitches based on various threshold values applied to different attributes.  The threshold values, though \textit{ad-hoc},  are chosen based on the simulation exercise.  These threshold choices vary based on the detector sensitivity during each run, the network combination, and the type of the transient signal.}

%comment:  Brief  update about  Machine learning method in GW
Machine learning (ML) techniques are the possible option to address such problems as they offer powerful tools for classification between signal and noisy transients \cite{noise_cl1,noise_cl2,noise_cl3,noise_cl4,noise_cl5,noise_cl6,noise_cl7,noise_cl8,noise_cl9,noise_cl10}. More generically, ML ideas have the notable potential to improve the detection sensitivity of unmodeled GW signals \cite{gmm,cwb_ml1,cwb_ml2,Mishra2021, Cuoco_2020}. ML techniques have also been used to enhance searches and classification of GW signals from CCSN \cite{ml_ccsn}.
%ML method is furthermore employed for the search and classification of GW signals from CCSN
%\blue{Siong: We should add a placeholder reference for the XGBoost paper as well}

%comment: OUR WORK
\dl{In an earlier work \cite{gmm},  the authors had proposed an alternative ML based approach to address the \textit{ad-hoc} thresholding on the attribute set. In this Gaussian mixture model (GMM)-based
supervised machine learning approach, we modeled the cWB trigger attributes in the multidimensional attribute space. Thus, it provides an alternative approach to  thresholds applied on multidimensional space to the detection problem under the scalar log-likelihood ratio. In this work, we continue to use the Gaussian mixture models in obtaining the model for the attributes in the multidimensional parameter set.  Here, we explore this approach by carefully choosing the attribute set  and investigating the dependence on the search sensitivity with the appropriate usage of the re-parametrization approach. We present the search sensitivity for the generic short-duration gravitational-wave transients using the data from the O3a run with the low-frequency (16–1024 Hz) analysis.}

\dl{The paper is organized as follows. Section \ref{Sec:methodology} describes the unmodeled search algorithm cWB and Gaussian mixture modeling for cWB triggers. Section \ref{Sec:analysis} discusses the dataset used to study the GMM approach to all-sky search, Gaussian mixture model generation, and sensitivity improvement to generic signal morphologies and core-collapse supernova waveforms. Section  \ref{Sec:results} discusses the GMM method results on cWB triggers of coincident events from the O3a run of Advanced LIGO detectors. Finally, in Section \ref{Sec:discussions}, we summarize the significance of the GMM method on an all-sky search for short-duration transient signals using a minimal model approach.}

\section{METHODOLOGY}

\label{Sec:methodology}

\subsection{Coherent WaveBurst algorithm}

%% Comment:  Brief introduction about cWB and define triggers in cWB. 

The cWB algorithm is a morphology-independent,  multidetector GW signal detection algorithm that coherently maps the multidetector's data into the multiresolution time-frequency scale domain using the Wilson-Daubechiers-Meyer wavelet transformation \cite{wdm}. It is based on the constraint maximum-likelihood ratio approach applied to the strain data in the time-frequency domain. The clusters of excess energy pixels are selected above the noise level of the detector and labeled as triggers if they exceed the thresholds on coherent energy ($E_c$) and network correlation coefficient ($c_c$) \cite{cwb_Klimenko,cwb_2021}. For each trigger, the cWB computes a collection of attributes that characterize signal as well as noisy transients properties. 

We consider the following attribute sets for this study,  which are generic for the short-duration transient signals. The network coherent energy $E_{c}$, the effective correlated signal-to-noise $\eta_{c}$, the network correlation coefficients  $c_{c0}$ and $c_{c2}$, network energy short-duration $N_{ED}$, the ratio between the reconstructed energy and the total energy  $N_{\mathrm{norm}}$, the residual noise energy measure  $\chi^{2}$ and attributes pertaining to the noise vetoes like $Q_{\mathrm{veto}}$ and  $L_{\mathrm{veto}}$.  We list this attribute set  in Table \ref{tab:attributes} \cite{cwb_page}.  In standard cWB-based analysis, the threshold on the multidimensional attribute space is placed to distinguish the noise based triggers from the astrophysical GW event.

\begin{table*}[htp]
%\centering
\begin{tabular}{c c}%{@{\extracolsep{\fill}}cccr}
\hline
  \hline
 Attribute & Definition \\ [1ex] 
 \hline
 
$E_{c}$ &   Cross correlation of reconstructed waveforms  between the detector pairs.\\
$\eta_{c}$ &   $\eta_{c} = \sqrt{\frac{E_{c}}{(K-1)}}$, where $K$ is the number of detector.\\ 
$c_{c0}$ & $c_{c0}=\frac{E_{c}}{(|E_{c}|+E_{n})}$, where $E_{n}$  is the energy of the residual noise.\\
$c_{c2}$ & $c_{c2}=\frac{E_{c} \times c_{c0}}{(|E_{c}|+E_{n})}$\\
$N_{ED}$ & Energy disbalance of the trigger between the detectors.\\
$N_{\mathrm{norm}}$ & Ratio between the reconstructed energy and the total energy. \\
$\chi^{2}$ &  $\chi^{2} = \frac{E_{n}}{N}$,  where N is the number of independent wavelet amplitude of the event. \\ % Minimize the non-Gaussian noise.
$Q_{\mathrm{veto0}}$ & Energy distribution of an event over different time segments. \\
$Q_{\mathrm{veto1}}$ & An estimate of quality factor assuming the signal to be a CosGaussian  (Q factor).\\ %, to identify blip-type glitches
$L_{\mathrm{veto0}}$ & Central frequency of the reconstructed signal, to identify narrow band glitches.\\
$L_{\mathrm{veto1}}$ & Root mean square frequency of the reconstructed signal.\\
$L_{\mathrm{veto2}}$ & Energy ratio between pixel energy and total energy of the event.\\[1ex] 
\hline

\end{tabular}
\caption{Summary of the selected attributes associated with the short-duration burst transients in cWB and used for the GMM-based analysis.}
\label{tab:attributes}
\end{table*}

%%  Comment: In this section we briefly explain the methodology of GMM. 
\subsection{Gaussian mixture modeling in postproduction }
 \dl{In \cite{gmm}, we proposed the Gaussian mixture modeling approach \cite{bishop:2006} to construct two distinct GMM models in the multidimensional trigger attribute set for astrophysical GW signals and noise glitches. We  used the Bayesian information criterion (BIC) method to obtain the optimum number of  Gaussians and the GMM model parameters (mean, covariance and weights).  For each model (signal and noise), we had obtained the maximum log-likelihood statistics $W = \ln(\hat{\mathcal{L}} )|_{\hat{K}}$, where $\hat{\mathcal{L}}$ is the maximum value of the likelihood function with the optimum number of Gaussians $\hat{K}$ \cite{gmm,bic_78,bic}. Then, we defined the detection statistic, $T$,  as 
\begin{align}
\label{T}
T = W_{\text{s}}- W_{\text{n}}\,.
\end{align}
where the subscript ${\mathrm s}$ and ${\mathrm n}$ stand for signal and noise. We had used this quantity as a ranking statistic for each cWB trigger instead of applying threshold values on the multidimensional  attribute set.  We suggest the reader to follow \cite{gmm} for more details.}

%As an example,  we show in Fig.\ref{fig:corner_plot}(top) the distribution of cWB trigger attributes for the signal model initially chosen for GMM analysis.  Further,
\dl{A well behaved multidimensional Gaussian distribution is preferred to obtain GMM. When the distribution of a selected subset of attributes is not well behaved (highly peaked), the GMM overestimates the required number of Gaussians. Thus, we either precondition the data by reparametrizing some of the cWB attributes or combining two attributes in a single attribute. } 

We reparameterize the subset of attributes  $\eta_{c}$, $c_{c0}$, $c_{c2}$, $N_{ED}$, $E_{c}$, $Q_{\mathrm{veto}}$ and $L_{\mathrm{veto}}$, which shows the non-Gaussian distribution by either applying a logarithmic or inverse sigmoid transformation. Additionally, we define a new parameter $L_{\mathrm{ratio}}: L_{\mathrm{veto1}}/L_{\mathrm{veto0}}$ instead of using $L_{\mathrm{veto0}}$ and $L_{\mathrm{veto1}}$ as two different attributes. The distribution of attributes after the reparameterization is well behaved and better conditioned than before reparametrization.  This shows a direct implication on the number of Gaussians in the GMM model.  For example,  for the data set considered here,  the number of optimum Gaussians in the original attribute set for the signal and noise model is 113 and 115, whereas the optimum number of Gaussians is reduced to 90 and 82 for the reparametrized attribute set.  More details about the dataset is in Sec.  \ref{Sec:analysis}.
More details on reparametrization can be found in Appendix.

\section{O3\lowercase{a} ANALYSIS WITH GMM}

\label{Sec:analysis}

%\subsection{Playground data and simulation details}

%%  Comment: Explain about O3A data (background and foreground data)
\dl{This study uses publicly available O3a data \cite{GWOSC_O3a,GWOSC} from the LIGO Hanford and LIGO Livingston (HL) network in the low-frequency range (16-1024 Hz)\footnote{As we focus on maximization of the detection efficiency,  we consider the HL network for this study rather than HLV because of the similar reasoning mentioned in \cite{allskyo3}.}. The total amount of coincidence data between the two detectors used for this analysis is 104.9 days. The distribution of accidental triggers is calculated by time-shifting the data of one LIGO detector with respect to the other LIGO detector by an amount that breaks any correlation between the detectors for an actual signal. For the cWB O3a analysis, 1000 years of background data were generated.}

\begin{table}[htp]
%\centering
\begin{tabular}{c c c c}%{@{\extracolsep{\fill}}cccr}
\hline
  \hline
 Attribute & LF1 & LF2 & LF3 \\ [1ex] 
 \hline
$c_{c0}$ & > 0.8 &> 0.8 &> 0.8\\
$c_{c2}$ & > 0.8 &> 0.8 & > 0.8 \\

$N_{\mathrm{norm}}$ & > 2.5 & > 2.5 & > 2.5\\

$Q_{\mathrm{veto0}}$ & = 0 & $\neq  0$ & \\
$Q_{\mathrm{veto1}}$ & $\leq 3$ & $\leq 3$ & > 3 \\
\hline

\end{tabular}
\caption{Definitions of the search bins used for the O3a  low-frequency short-duration burst analysis. The different thresholds are applied to attributes $c_{c0}$, $c_{c2}$, $N_{\mathrm{norm}}$, $Q_{\mathrm{veto0}}$ and  $Q_{\mathrm{veto1}}$ to classify the triggers into different bins based on background trigger distribution.}
\label{tab:bin_thershold}
\end{table}

\subsection{The standard O3a cWB all-sky analysis}

%Comment: Explain about standard cWB approach to distinguish signal and noise
\dl{In standard cWB O3a analysis, triggers are required to pass frequencies above 24 Hz and high network correlation with $c_c$ above 0.8 for the HL detectors. These triggers are classified into three different bins  as LF1, LF2, and LF3 based on background trigger morphologies to isolate the triggers to a small part of the parameter space  \cite{allskyo3}.  LF1 contains a population of glitches dominated by very short and loud blip-type glitches, with negligible energy outside the single oscillation \cite{blip_glitch}. LF2 contains triggers resembling  blip glitches with Q factor $\leq$ 3, and LF3 contains remaining high Q factor triggers. Table \ref{tab:bin_thershold} lists the thresholds applied to cWB attributes for classification into different bins used in O3a analysis. After applying thresholds on the cWB attributes, the events are ranked based on inverse false alarm rate (iFAR). A trials factor of three is applied to iFAR values corresponding to three different background bins, and a threshold of 100 years on iFAR is used for the significant detection \cite{allskyo3}. By contrast, GMM based postprocessing does not require the creation of multiple analysis bins (see below).}

\subsection{GMM model generation}

For O3a GMM analysis, we choose most of the attributes required to characterize the generic short-duration transient signal from standard cWB (Table \ref{tab:attributes}) to generate the GMM model. 
We exclude the attributes reconstructed signal's central frequency, duration, and strain from this analysis, as these  attributes are strongly affected by the priors on the injection parameters for the foreground model  \cite{gmm,cwb_page}. Since we do not want the GMM to develop a bias towards the choice of the signal population used in the simulation.  For GMM analysis, we consider those triggers with $\eta_{c} > 5.5$ and $c_{c}>0.5$  for the HL network.  The cWB uses different thresholds on attributes to generate gravitational wave candidates and veto out the noise triggers in its standard configuration. 

We consider $70\%$ of noise background as well as simulated burst signals for training and the rest for testing the GMM method. We use the equal percentage of generic waveforms (Gaussian Pulse, sine-Gaussian wavelets, and white noise burst) as the training set aimed to improve the detection efficiency of all injected waveforms. We construct the GMM for the noise and signal triggers using this simulation \cite{gmm}. More specifics of the models can be found in Appendix.

\subsection{Data simulation}

%% Comment: Explain about injection signals used for signal model.
\dl{A set of generic short-duration burst waveforms (typically used in the all-sky short-duration bursts  search \cite{allskyo3}) are injected in the HL O3a noise to estimate the search sensitivity using the cWB plus GMM method. They are sine-Gaussian wavelets (SG), Gaussian pulses (GA), and band-limited white-noise bursts (WNB) signals injected over a range of amplitude expressed in terms of the root-sum-squared strain amplitude ($h_{\mathrm{rss}}$). The SG waveforms are characterized by mean frequency $f_{0}$ and quality factor $Q$,  the GA waveforms are described by the duration $\tau$, and the WNB signals are characterized by lower frequency bound $f_{\mathrm{low}}$,  frequency bandwidth $\Delta{f}$, and duration $\tau$.  The SG and GA signals are injected as $h_{\mathrm{rss}} = (\sqrt{3})^{N}5 \times 10^{-23} Hz^{-1/2}$ over a grid of maximum strain values with N ranges from 0 to 8,  and  WNB waveforms are injected uniformly as the square of the signal's distance. Linearly polarized signals are used  for GA, and sine-Gaussian wavelets use both elliptical (SGE) and linearly (SG) polarized waveforms \cite{allskyo3, allskysixth}. Elliptical waveforms are uniform in the cosine of the source inclination angle, which is defined as the angle between total angular momentum and the line of sight. The WNB represents isotropic emission at the source and carries an equal amplitude from both polarization of GW strain at the detector \cite{Sutton_wnb}. Table \ref{tab:waveforms} lists the simulated signals and their characteristic parameter values used in this analysis.}

\begin{table}[htb]
\begin{tabular}{c c c c}
\hline
\hline
\multicolumn{3}{c}{Sine-Gaussian Burst (SGW)}\\
\hline
No. &$f_0$ (Hz) & $Q$ &  \\
\hline
1 & 70 & 3 &  \\
2 &70 & 9 &  \\
3 &70 & 100 &  \\
%\hline
4 &100 & 9 &  \\
5 &153 & 9 & \\
%\hline
6 &235 & 3 &  \\
7 &235 & 9 &  \\
8 &235 & 100 &  \\
%\hline
9 &361 & 9 &  \\
10 &554 & 9 & \\
%\hline
11 &849 & 3 & \\
12 &849 & 9 & \\
13 & 849 & 100 & \\
\hline
%\end{tabular}
%\begin{tabular}{|c|c|c|}
\multicolumn{3}{c}{White-Noise Burst (WNB)}\\
\hline
 & $f_{\text{low}}$ (Hz) & $\Delta f$ (Hz) & $\tau$ (s)\\
\hline
14 & 150 & 100 & 0.1 \\
%\hline
15 & 300 & 100 & 0.1 \\
16 & 750 & 100 & 0.1 \\

%\hline
\hline
%\end{tabular}
%\begin{tabular}{|c|}
%\hline
\multicolumn{3}{c}{Gaussian Pulse (GP)}\\
\hline
 &  & & $\tau$ (s) \\
\hline
17 & & & 0.1 \\
18 &  & & 1 \\
%\hline
19 & & & 2.5 \\  
20 & & & 4 \\
\hline
\hline
\end{tabular}
\caption{List of generic burst waveforms and their characteristic parameters.}\label{tab:waveforms}
\end{table}

%% Comment: Explain about Supernova waveforms.
\dl{The low-frequency,  generic all-sky short-duration burst search  targets the frequency range of GW burst from most of the core-collapse supernovae (CCSN). To demonstrate the  robustness of the cWB plus GMM approach against different signal morphologies, here,  we analyze the search sensitivity to CCSN  waveforms used in all-sky short-duration transient analyses during O3a  \cite{allskyo3}. The  waveforms are generated  from five different  3D CCSN simulations  models like $\mathrm{s18}$  \cite{s18},  $\mathrm{m20}$ \cite{m20},  $\mathrm{s9}$ \cite{s9}, $\mathrm{m39}$ \cite{m39}, and $\mathrm{35OC}$ \cite{35OC}. The waveforms are distributed uniformly in distance  with maximum distance for $\mathrm{s18}$, $\mathrm{m20}$, $\mathrm{s9}$, $\mathrm{m39}$, and  $\mathrm{35OC}$ CCSN models set to 25 kpc, 5kpc, 5kpc, 70kpc, and 70 kpc, respectively \cite{allskyo3}.}

% Sensitivity improvement to generic signal morphologies (Explain about ROC curve)
\subsection{ Sensitivity improvement for generic signal morphologies}

\begin{figure*}[htp] %htp
  %\centering
    {\includegraphics[width=1\linewidth,height=0.55\textheight]{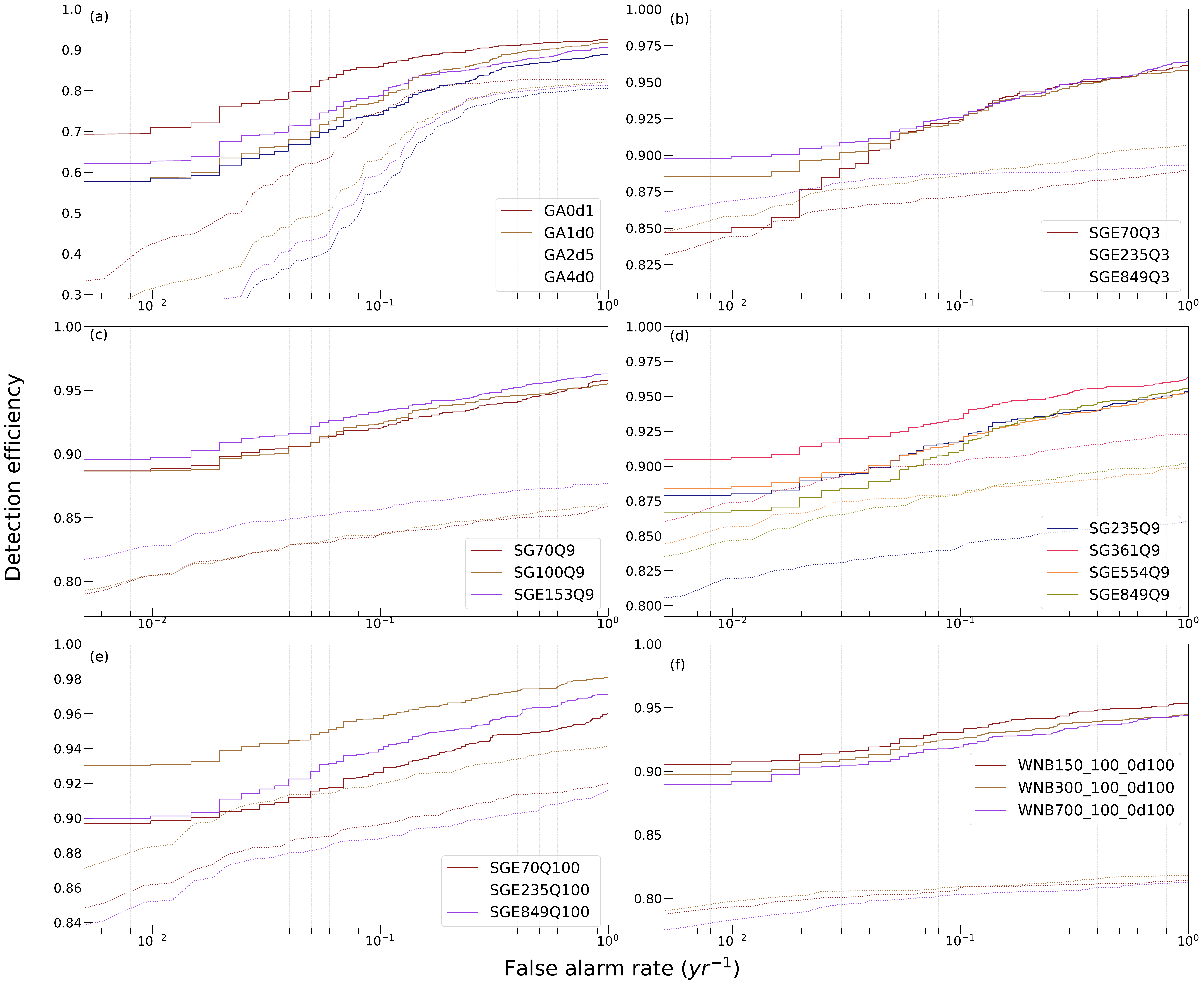}} %roc_curve.pdf
\caption{ROC curves for the cWB plus GMM search (solid lines)
compared to the standard cWB search (dotted lines) for the simulated waveforms. Panel (a) shows GA class waveforms with duration 0.1, 1, 2.5 and, 4 s. (b) shows elliptically polarized SG waveforms with a quality factor $Q=3$ and frequency at 70, 235, and 849 Hz. Panel (c) and (d) show the SG waveforms with $Q=9$ and 70-900 Hz frequency range. (e) shows SG with $Q=100$ and frequency at 70, 235, and 849 Hz. (e) WNB is characterized by $f_{\mathrm{low}}$ at 150, 300, and 750 Hz with frequency bandwidth and duration at 100 and 0.1 s.}
\label{fig:roc_curve}
\end{figure*}

We use the receiver-operating-characteristic (ROC) curve to characterize the sensitivity improvement. The ROC curve of the standard cWB and the cWB plus GMM relates the detection efficiency against the false alarm rate (FAR) is shown in Fig. \ref{fig:roc_curve}.  The detection efficiency is the fraction of the recovered injections by a given algorithm.  The standard cWB uses $\eta_c$, and cWB plus GMM uses T as a detection statistic.  We show that the cWB plus GMM algorithm significantly enhances the detection efficiency for all the 20 waveforms compared to the standard cWB. 

\begin{figure*}[t!]
  \centering
    \scalebox{0.235}[0.19]{\includegraphics{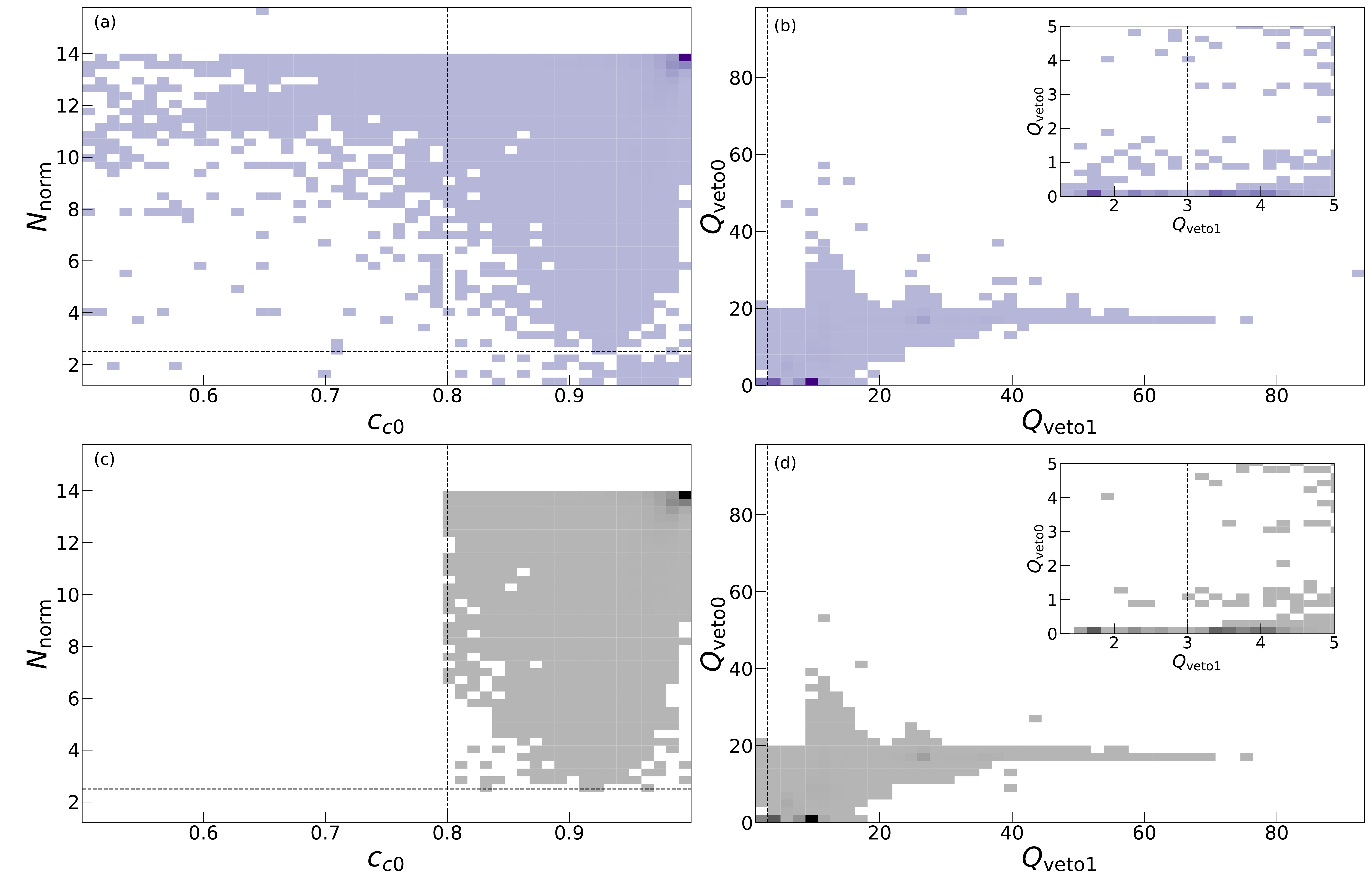}}
  \caption{The two-dimensional distribution of signal trigger attributes $c_{c0}$ \textit{vs} $N_{\mathrm{norm}}$ ((a) and (c)) and $Q_{\mathrm{veto1}}$ \textit{vs} $Q_{\mathrm{veto0}}$ ((b) and (d)) for triggers recovered by cWB plus GMM  (purple) and standard cWB (gray) at iFAR $\geq$ 100 years.  The inset shows a zoom-in of the 2D distribution of $Q_{\mathrm{veto1}}$ \textit{vs} $Q_{\mathrm{veto0}}$. The black dashed lines show the postproduction threshold applied to these attributes for the standard cWB all-sky search, which classify the events into various bins as shown in Table \ref{tab:bin_thershold}.}
  \label{Fig.2dhist}
\end{figure*}

In order to understand this improvement,  we show the two-dimensional distribution of attributes $c_{c0}$, $N_{\mathrm{norm}}$, $Q_{\mathrm{veto0}}$ and $Q_{\mathrm{veto1}}$ of the signal triggers with iFAR $\geq$ 100 years in cWB plus GMM (by purple color) and standard cWB (by gray color) analysis in Fig. \ref{Fig.2dhist}. From panels (a) and (c), it is evident that with GMM, we have improved the signal/noise classification ability that we can now recover events below $c_{c0}$ of 0.8.

The most significant improvement after using cWB plus GMM is shown in the recovery of GA morphology (see Fig. \ref{fig:roc_curve}(a)).  The GA signal mostly resembles blip glitches and lies mainly in bin LF1.  During the O3 run, the search sensitivity of standard cWB to the GA waveforms has worsened compared to its sensitivity O2.  This is primarily due to an abundance of blip glitches  that resemble GA signals. However, with GMM based postproduction approach appears to have modeled them well, which is also evident in the recovery of GA with $Q_{\mathrm{veto1}} <3 $ (in Fig \ref{Fig.2dhist} panels (b) and (d)) those otherwise would get classified as bin LF1 events in the standard cWB.  In standard cWB analysis, detection efficiency for triggers falls in each bin estimates separately. However, the signal waveforms in bin LF1 are mostly buried in background glitches.  As a result, cWB plus GMM has shown an enhanced sensitivity for Gaussian pulses with an improvement of greater than $20\%$ at an iFAR of 100 years. The signal waveforms lying in bin LF1 show comparatively lower significance to other morphologies in both cWB plus GMM and standard cWB method at the given signal strength because of excess glitches.

The cWB plus GMM approach marginally improves the SG class of injections  (Fig. \ref{fig:roc_curve}(b)-(e)). This indicates that our approach can recover other burst morphologies considerably better than standard cWB. For WNB,  the cWB plus GMM approach consistently improves the search sensitivity with respect to standard cWB (Fig. \ref{fig:roc_curve}(f)). WNB falls in the  bin LF3,  the cleanest part of the background distribution and cWB plus GMM has shown improvement of around $10\%$ throughout  the iFAR threshold.

It should also be noted that categorizing the events into various bins in standard cWB analysis iFAR is penalized by a trials factor of three. Whereas in cWB plus GMM based analysis, there is no such bin classification.  This can also be attributed to some improvement in detection efficiency.

\subsection{Robustness test to CCSN}

\begin{figure}[H]
  %\centering
    {\includegraphics[width=1\linewidth,height=0.193\textheight]{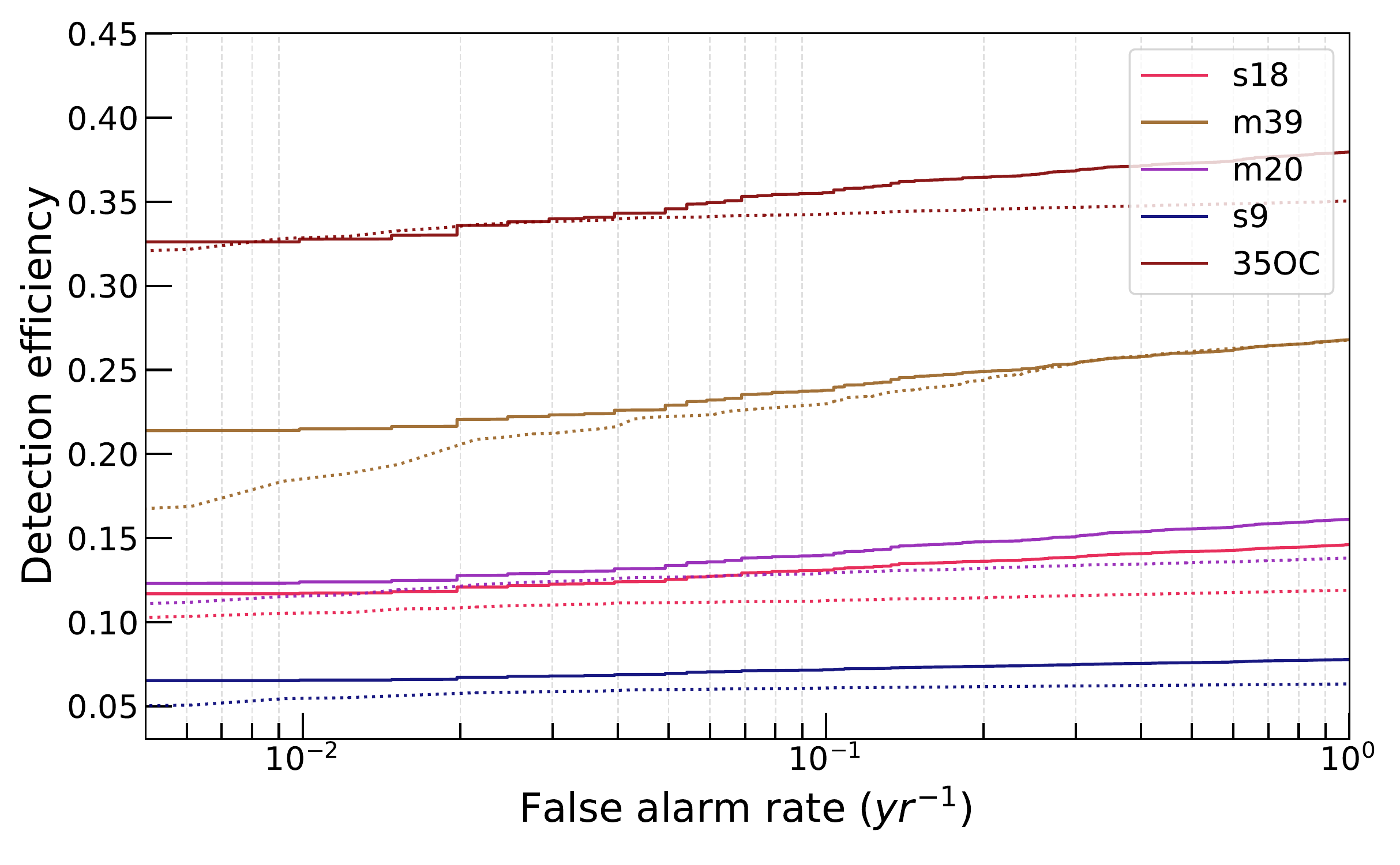}}
\caption{ROC curves for CCSN waveforms. As before, the cWB plus GMM search ROC curves are represented by solid lines, while the standard cWB search ROC curves are plotted as dotted lines.}
\label{fig:roc_sn}
\end{figure}

To test the robustness against the different morphologies of waveforms and distribution, we applied the GMM model on the core-collapse supernovae (CCSN) injections used in  \cite{allskyo3}  that are not included in the training data set. Fig. \ref{fig:roc_sn} shows the ROC curves for injections with different CCSN waveforms. The  models $\mathrm{35OC}$ and $\mathrm{m39}$  describe the explosion driven by rapid rotation of  massive progenitor star with  zero age main sequence (ZAMS) mass of $35 M_{\odot}$ and $39 M_{\odot}$, which creates strong GW emission \cite{allskyo3}. Both standard cWB and cWB plus GMM show comparable high efficiency to CCSN model $\mathrm{35OC}$ at low FAR because GW signal occurs frequency above 100 Hz, which is optimally sensitive to the cWB low-frequency search method. However, the CCSN model $\mathrm{m39}$,  cWB plus GMM enhances the detection efficiency compared to the standard cWB. Since  GW amplitude peaks at a frequency $\sim$ 750 Hz, at the edge of the postproduction threshold  applied for standard cWB. The  standard cWB and cWB plus GMM give comparatively low efficiency for all the other models such as $\mathrm{m20}$, $\mathrm{s18}$ and $\mathrm{s9}$, which describes GW emission from nonrotating progenitors with ZAMS mass ranges from $9 M_{\odot}$ to $20 M_{\odot}$. Among them,  cWB plus GMM performs better than standard cWB at all FAR threshold values.  The ML based cWM plus GMM method does not focus on one class but shows good improvement for a broad class of waveforms at detectable iFAR levels. %\sout{for CCSN model $m39$ cWB plus GMM enhance the detection performance at low FAR. The other three CCSN models $m20$, $s18$ and $s9$ standard cWB and cWB plus GMM, achieve low efficiencies at given FAR with a raise in detection performance compared to the standard-cWB approach.}

\section{SEARCH RESULTS}
\label{Sec:results}

\begin{figure}[H]
  %\centering
    {\includegraphics[width=1.0\linewidth,height=0.2\textheight]{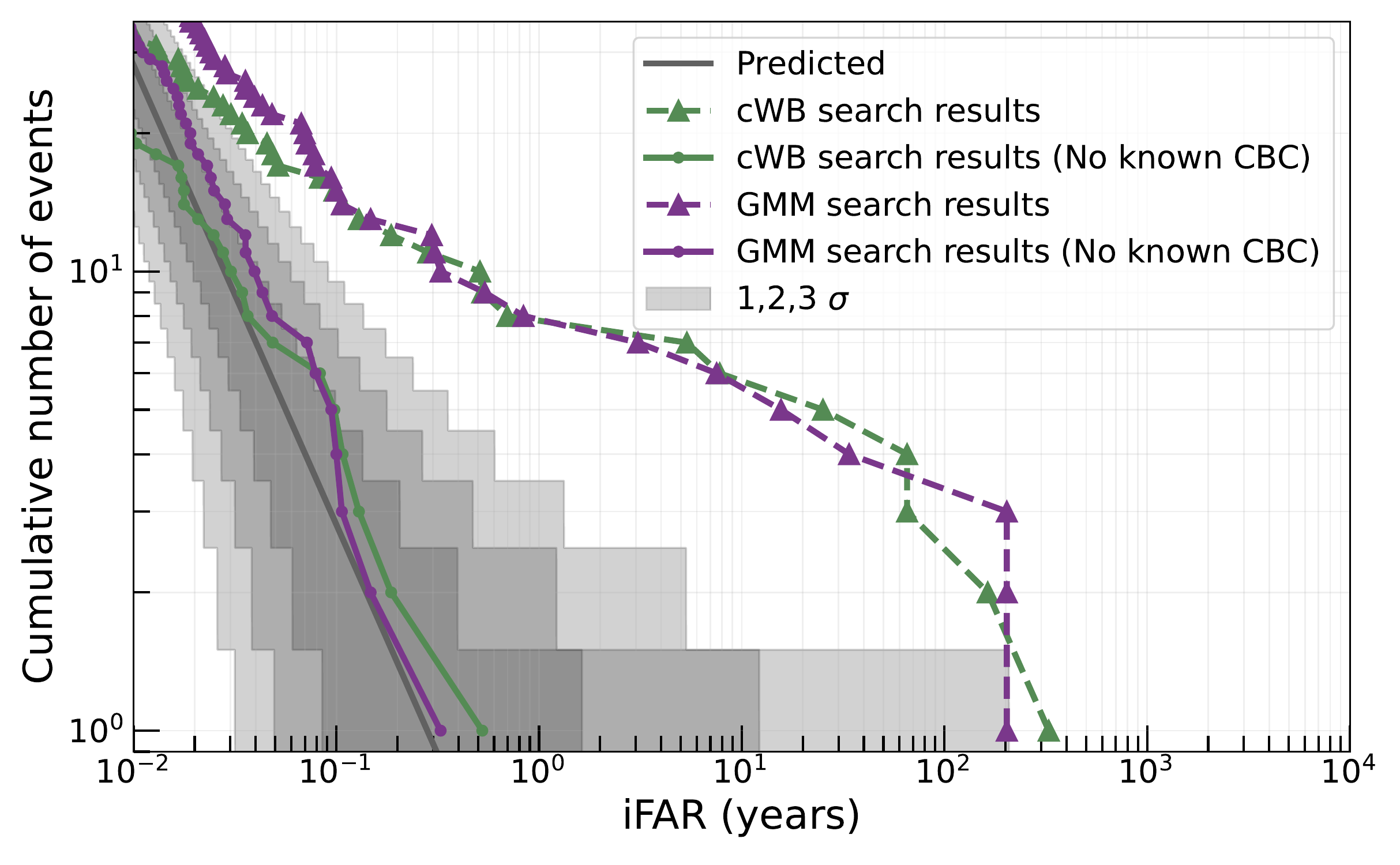}}
\caption{The cWB plus GMM (purple) and standard cWB  (green) observed the cumulative number of events versus iFAR in the low-frequency range. The iFAR distribution for all coincident data is plotted using triangular markers, while the line with circular markers shows the distribution after the removal of known CBC events. The expected mean value of background shows as a solid gray line with 1, 2 and 3$\sigma$ Poisson uncertainty (shaded regions).}
\label{fig:zerolag}
\end{figure}

%Comment : Search result for Foreground data. Mention about  non CBC foreground events and CBC events which are reported in catalog paper found by GMM and cWB all sky search 
\dl{Following the sensitivity study,  we applied the cWB plus GMM on the O3a data from LIGO Hanford and LIGO Livingston. 
The maximum background for cWB plus GMM analyses is 200 years since we used 20\% of the total O3a background as test data. 
The results of this search from the O3a low-frequency region are shown in Fig. \ref{fig:zerolag}.  We plot the cumulative number of recovered events by standard cWB (by green color) and cWB plus GMM (by purple color) in terms of their iFAR values.}

\dl{The cWB plus GMM method recovers all CBC events identified by the targeted cWB search for CBC in GWTC-2 and GWTC-2.1~\cite{gwtc2, gwtc2_1,3-OGC}. No new candidates were found with high statistical significance in this search. The most significant non-CBC event observed at UTC 2019-09-30 23:46:52 with an iFAR of 0.3 years in the cWB plus GMM method shows 0.01 years of iFAR in the standard cWB search. The second most significant event at UTC 2019-05-11 04:12:15 was with an iFAR of 0.15 years in cWB plus GMM, which had shown iFAR of 0.002 years in standard cWB.  We note that the two loudest non-CBC events in standard cWB all-sky search are at UTC 2019-09-28 02:11:45 and UTC 2019-08-04 08:35:43, with an iFAR of 0.53 years, and 0.19 years~\cite{allskyo3}, showed less significance in the cWB plus GMM search with an iFAR of 0.006 and 0.05, respectively.  After excluding the CBC signals, the distribution of coincident events is close to the predicted background for the analyzed time.} 

Table~\ref{tab:cbc} lists the CBC events from GWTC-2 and GWTC-2.1 that are also observed by standard cWB all-sky burst search \cite{allskyo3}. We list their attribute values,  $T$ statistic value, total source mass,  and the corresponding iFAR values for both standard all-sky burst cWB search~\cite{allskyo3} and cWB plus GMM search~\footnote{The significance values for standard cWB shown in this table are different from those listed in GWTC-2 and GWTC-2.1 because the significance estimates in these publications are based on a version of cWB optimized for CBC events. Since the cWB plus GMM approach is applied in the all-sky search context, we find it fairer to compare the event significance obtained for all-sky searches.}. We observe that the cWB plus GMM method is less sensitive to low mass BBH systems (with total source mass < 60 $M_{\odot}$) since these events are more significant in the standard cWB all-sky search. The exception is GW190412,  which has significantly asymmetric component masses and shows the contribution from higher modes\footnote{The component masses for GW190412 system are $m_{1} = 30.1^{+4.6}_{-5.3} M_{\odot}$ and $m_{2}=8.3^{+1.6}_{-0.9} M_{\odot}$.} \cite{GW190412}.  The IMBH system GW190521 \cite{GW190521} of total mass $163 M_{\odot}$ was observed with high significance in cWB plus GMM with an iFAR of greater than 200 years, which shows an iFAR for 65 years in all-sky burst cWB search \cite{allskyo3} \footnote{GW190521 reported with an iFAR greater than 4900 years using targeted CBC searches \cite{GW190521}.}.  A similar trend  is observed for other events in Table~\ref{tab:cbc}, with total source mass $M > 100 M_{\odot}$. The CBC event count acts as a verification step just to review the sensitivity of cWB plus GMM search against standard cWB all-sky search.

\begin{table*}[htp]
\centering
\scriptsize
\begin{tabular}{c c c c c c c c c c}%{@{\extracolsep{\fill}}cccr}
\hline
 \hline
 & &  &   &  &  & & & \multicolumn{2}{c}{iFAR in years} \\
 \hline
 Event & $\eta_{c}$ & $Q_{\mathrm{veto}}$ & $c_{c0}$ & $\chi^{2}$ & $N_{\mathrm{norm}}$ & $T$ & $M(M_{\odot})$ & cWB+GMM & cWB\\ [1ex] 
 \hline\hline 
 GW190408\_181802 & 8.59 & 0.92 & 0.96 &0.13 & 5.09& -0.41& $43.0^{+4.2}_{-3.0}$  &0.30 &25.14 \\ 
 GW190412 & 11.69 & 4.16 & 0.95 &0.06 &5.4 & 13.21& $38.4^{+3.8}_{-3.7}$ &15.62&14.86  \\
 GW190421\_213856 & 6.46 & 0.31 & 0.97 & -0.07 & 4.41 &-0.38 & $72.9^{+13.4}_{-9.2}$ & 0.30&0.04 \\
 GW190426\_190642 & 5.52 & 0.45 & 0.88 & 0.08 & 4.07 & -4.85 & $184.4^{+41.7}_{-36.6}$ & 0.02 & 0.01 \\
 GW190503\_185404 & 7.34 & 0.34 & 0.93 &-0.02 &4.76 &1.65 & $71.7^{+9.4}_{-8.3}$ & 0.84& 0.70\\
 GW190513\_205428 & 7.05 & 1.67 & 0.86 & 0.15 & 3.77 &-2.99 & $53.9^{+8.6}_{-5.9}$ & 0.07& 0.28\\
 GW190517\_055101 & 6.08 & 0.19 & 0.88 & -0.15 & 3.05 & -2.79 & $63.5^{+9.6}_{-9.6}$ &0.08 &0.01 \\
 GW190519\_153544 & 10.13 & 0.53 & 0.89 & 0.01 & 7.63 &18.04 & $106.6^{+13.5}_{-14.8}$ & 33.83 &7.78 \\
 GW190521 & 9.24 & 0.60 & 0.92 &-0.16 & 10.53 &32.45 & $163.9^{+39.2}_{-23.5}$ & $>200 $ &65.38 \\
 GW190521\_074359 & 14.19 & 0.56 & 0.96 &-0.08 &8.44 &72.77 & $74.7^{+7.0}_{-4.8}$ & $>200$ & 326.88\\
 GW190602\_175927 & 7.25 & 0.43 & 0.95 & -0.13 &6.5 & 0.73 &  $116.3^{+19.0}_{-15.6}$ &0.54&0.51 \\
 GW190706\_222641 & 9.29 & 0.79 & 0.83 & -0.10 &7.36 &24.93 & $104.1^{+20.2}_{-13.9}$ & $>200$ &65.38 \\
 GW190727\_060333 & 5.86 & 0.35 & 0.96 &  0.17 &4.96 & -2.94 & $67.1^{+11.7}_{-8.0}$ &0.07 & 0.006\\
 GW190728\_064510 & 6.50 & 3.94 & 0.87 &-0.13 & 2.55& -4.93 & $20.6^{+4.5}_{-1.3}$ &0.02 &0.051 \\
 GW190828\_063405 & 10.27 & 0.84 & 0.82 &0.10 & 5.01&8.78 & $58.0^{+7.7}_{-4.8}$ & 7.52&163.44 \\
 GW190915\_235702 & 8.07 & 0.42 & 0.95 &0.06 &4.29 &5.29 & $59.9^{+7.5}_{-6.4}$ & 3.07&5.36 \\
 GW190929\_012149 & 5.97 &  0.22 &  0.85 &0.103 &3.44 &-6.20 & $104.3^{+34.9}_{-25.2}$ & 0.01&0.009 \\[1ex] 
 \hline
\end{tabular}
\caption{Table of CBC candidate events identified by cWB in GWTC-2 and GWTC-2.1. The columns show the effective correlated signal-to-noise $\eta_{c}$, energy distribution of the event over different time segments  $Q_{\mathrm{veto0}}$, network correlation coefficients  $C_{c0}$, residual noise energy measure  $\chi^{2}$, the ratio between the reconstructed energy and the total energy  $N_{\mathrm{norm}}$  and source total mass $M$ reported in GWTC-2 \cite{gwtc2} and GWTC-2.1 \cite{gwtc2_1}. The two rightmost columns report event significances for the cWB plus GMM search and the standard cWB all-sky search.}
\label{tab:cbc}
\end{table*}

\section{DISCUSSIONS}
\label{Sec:discussions}

% Conculsion about the current work
\dl{This paper provides a robust method of direct usage of a supervised machine learning approach at the postproduction stage of a well-established model-independent cWB detection algorithm. The work employs a Gaussian mixture modeling approach to model cWB triggers (signal as well as noise) in a multidimensional attribute set. The GMM modeling helps to classify background glitches from the transient GW signals in a log likelihood-based test statistic. While this approach was initially conceptualized in \cite{gmm} here, we extend it further in conditioning the attribute set by first taking an appropriate subset targeted for a given signal class. Then introduce appropriate reparametrization of the attributes ensuring the well-behaved distributions for the construction of GMM.}

\dl{We use the cWB triggers set from the all-sky short-duration burst search during O3a coincident data from the two Advanced LIGO detectors. We consider the signals from generic morphology and specific CCSN models used in \cite{allskyo3} for sensitivity study. For generic morphology, there is a definite improvement in sensitivity for a fixed value of false alarm rate for all waveforms using the cWB plus GMM method over the standard cWB. We observed notable improvement for the Gaussian pulse, which falls in the bin LF1 containing a very short and loud (blip-type) population of glitches. Sensitivity improvement to CCSN models clearly demonstrates the robustness of the cWB plus GMM method against the variation in morphology as well as distribution of waveforms as CCSN waveforms were not part of the training set. }

\dl{The application of cWB plus GMM on the coincident data search is consistent with the all-sky short-duration burst results using the standard cWB search as published in  \cite{allskyo3}.  No additional, significant events are observed.  The search recovered the GWs signals from the BBH merger as found by the targeted search for CBC signals.  We observed higher significance for massive CBC events compared to that reported in the all-sky burst \cite{allskyo3}. This clearly shows the promise of the cWB plus GMM approach at the postproduction stage of the burst algorithm as it gives a better handle in distinguishing the short-duration bursts from that of the short-duration noisy glitches. We plan to further extend this work to Generalized mixture modeling \cite{generalized}, which allows us to model more complex distributions for the fourth and fifth observing runs of ground-based GW detectors~\cite{o4_o5}.}

\begin{acknowledgments}
The authors thank Shubhanshu Tiwari for valuable discussion and suggestions. D.L. acknowledges support from Swiss National Science Foundation  (SNSF) grant number 200020-182047.  A.P. acknowledges the SERB Matrics grant MTR/2019/001096 and SERB-Power-fellowship grant SPF/2021/000036 of Department of Science and Technology,  India for support. The authors are thankful to LIGO-Virgo-Kagra collaboration for provisions of the cWB attribute data used for the all-sky short-duration O3a analysis. The authors are grateful for the computational resources and data provided by the LIGO Laboratory and supported by National Science Foundation Grants No.PHY-0757058 and No.PHY-0823459. C.M. and I.S.H. are supported by the Science and Technology Research Council (grant No. ST/V005634/1) and the European Cooperation in Science and Technology (COST) action CA17137. The open data is available in the Gravitational Wave Open Science Center \cite{gwopen}, a service of LIGO Laboratory, the LIGO Scientific Collaboration and the Virgo Collaboration. The authors also acknowledge the use of the LDG clusters for computational/numerical work.
\end{acknowledgments} 

%\appendix

\renewcommand{\thesubsection}{\Alph{subsection}}

\section*{APPENDIX: REPARAMETRIZATION OF ATTRIBUTES}
\label{app:reparameterization}

%\begin{widetext}
\begin{table}[!htp]
\caption{Details of reparametrization.}
\label{tab:repar}
\begin{subtable}{0.45\textwidth}{
\centering
%\caption{(a)}
\begin{tabular}{c c}
%{@{\extracolsep{\fill}}cccr}
\hline
\hline
Original  attribute set & Reparametrized  attribute set \\ [1ex] 
 \hline 
$\eta_{c}$ & $log_{10}(\eta_{c})$\\
$c_{c0}$ & $logit(c_{c0})$\\
$c_{c2}$ & $logit(c_{c2})$\\
$N_{ED}$ & $log_{10}(N_{ED} + 10^{3})$\\
$E_{c}$ & $log_{10}(E_{c})$\\
 $N_{\mathrm{norm}}$ &  $N_{\mathrm{norm}}$\\
 $\chi^{2}$ &  $\chi^{2}$\\
 $Q_{\mathrm{veto0}}$ & $log_{10}(Q_{\mathrm{veto0}} + 1)$\\
 $Q_{\mathrm{veto1}}$ & $log_{10}(Q_{\mathrm{veto1}})$\\
 $L_{\mathrm{ratio}}$ & $logit(L_{\mathrm{ratio}})$\\
$L_{\mathrm{veto2}}$ & $logit(L_{\mathrm{veto2}} \times 0.99)$\\[1ex] 

\end{tabular}}
%\caption{Re-parameterization of cWB trigger attributes used for this study which, are generic for the short-duration transient signals.}
%\label{tab:repar}}
\end{subtable}
%\hspace{0.05in}
%\begin{table*}[h]
%\scriptsize
\vspace{0.4in}
\begin{subtable}{0.45\textwidth}{
\centering
%\caption{(b)}
\begin{tabular}{c c c}%{@{\extracolsep{\fill}}cccr}
  \hline
  \hline
  & \multicolumn{2}{c}{Number of optimum Gaussians in GMM} \\
  \hline
Model & Original attribute set & Reparametrized attribute set \\  [1ex]
 \hline
 Signal & 113 & 90\\
 Noise & 115 & 82 \\[1ex] 
\hline
\end{tabular}}
%\end{center}
%\caption{Effect of reparametrization}
%\label{tab:gaussians}}
\end{subtable}
\end{table}

\begin{figure*}
\centering     %%% not \center
\includegraphics[width=0.9\textwidth,height=0.68\linewidth]{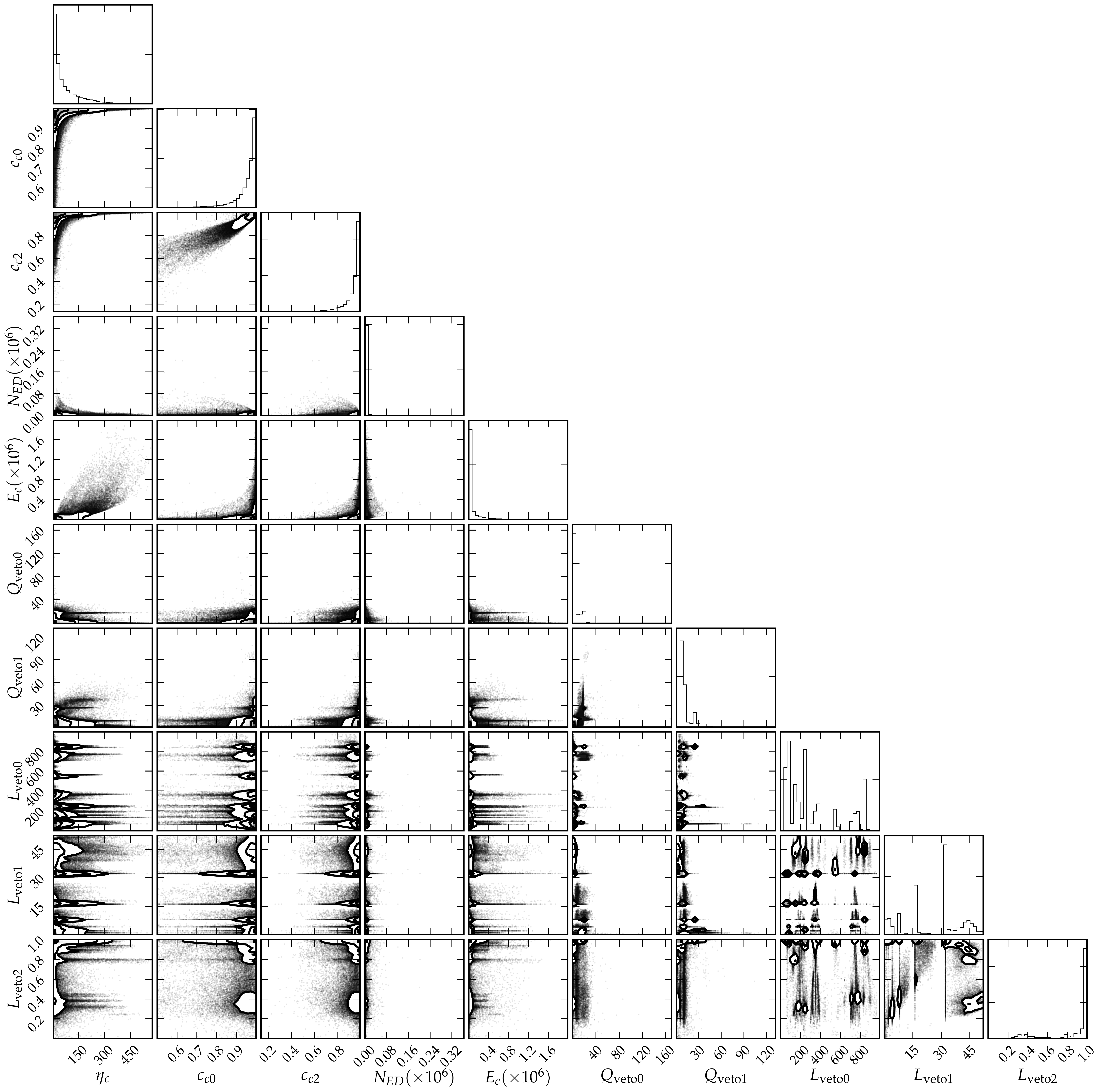}  %[height=12cm, width=14.5cm] : full size %width=10.5 cm, height = 10.5 cm
\hspace{\fill}
\includegraphics[width=0.9\textwidth,height=0.68\linewidth]{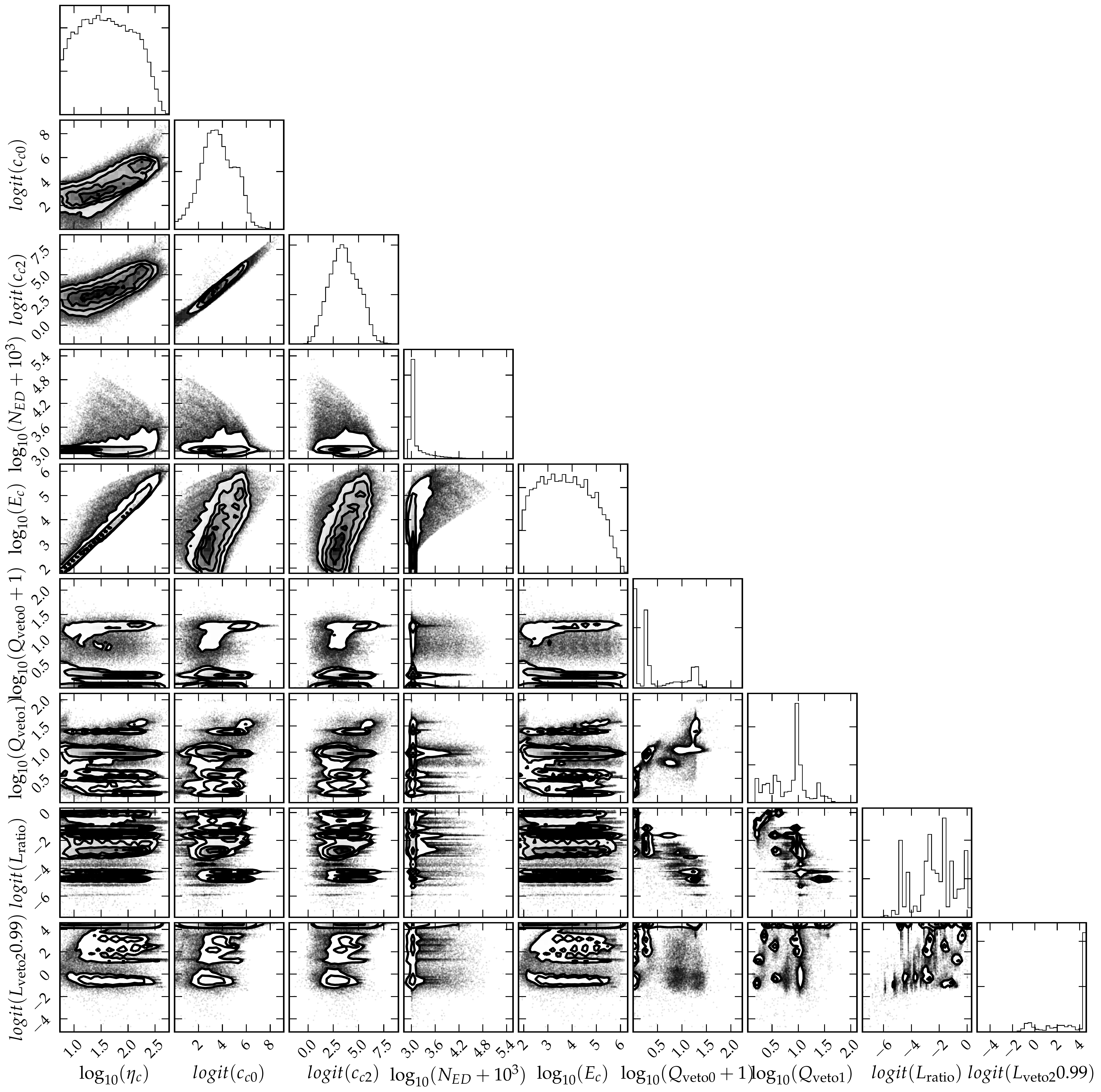}
\caption{The corner plot showing the one and two dimensional distribution of the reparametrized cWB trigger attribute for the signals before (top) and after (bottom) the reparameterization.} %We combined $L_{veto0}$ and $L_{veto1}$ as  $L_{ratio}: L_{veto1}/L_{veto0}$.}
\label{fig:corner_plot}
\end{figure*}

Given the cWB trigger attributes $E_{c}$, $\eta_{c}$, $c_{c0}$, $c_{c2}$, $N_{ED}$,  $N_{\mathrm{norm}}$,  $\chi^{2}$, $Q_{\mathrm{veto0}}$ ,$Q_{\mathrm{veto1}}$, $L_{\mathrm{veto0}}$, $L_{\mathrm{veto1}}$, and  $L_{\mathrm{veto2}}$. We reparametrize  some of the attributes which do not follow a well behaved Gaussian distribution. In Table \ref{tab:repar}, we show the reparametrized attributes along with their original form. We further show the reduction in the number of Gaussians after the reparameterization of the attributes. The distribution of reparametrized cWB trigger attributes for the signal initially chosen for GMM analysis and after the reparameterization is shown in Fig. \ref{fig:corner_plot}.

%\newpage
\clearpage 

\bibliographystyle{apsrev4-1}
% Produces the bibliography via BibTeX.
\bibliography{ref}

\begin{thebibliography}{59}%
\makeatletter
\providecommand \@ifxundefined [1]{%
 \@ifx{#1\undefined}
}%
\providecommand \@ifnum [1]{%
 \ifnum #1\expandafter \@firstoftwo
 \else \expandafter \@secondoftwo
 \fi
}%
\providecommand \@ifx [1]{%
 \ifx #1\expandafter \@firstoftwo
 \else \expandafter \@secondoftwo
 \fi
}%
\providecommand \natexlab [1]{#1}%
\providecommand \enquote  [1]{``#1''}%
\providecommand \bibnamefont  [1]{#1}%
\providecommand \bibfnamefont [1]{#1}%
\providecommand \citenamefont [1]{#1}%
\providecommand \href@noop [0]{\@secondoftwo}%
\providecommand \href [0]{\begingroup \@sanitize@url \@href}%
\providecommand \@href[1]{\@@startlink{#1}\@@href}%
\providecommand \@@href[1]{\endgroup#1\@@endlink}%
\providecommand \@sanitize@url [0]{\catcode `\\12\catcode `\$12\catcode
  `\&12\catcode `\#12\catcode `\^12\catcode `\_12\catcode `\%12\relax}%
\providecommand \@@startlink[1]{}%
\providecommand \@@endlink[0]{}%
\providecommand \url  [0]{\begingroup\@sanitize@url \@url }%
\providecommand \@url [1]{\endgroup\@href {#1}{\urlprefix }}%
\providecommand \urlprefix  [0]{URL }%
\providecommand \Eprint [0]{\href }%
\providecommand \doibase [0]{http://dx.doi.org/}%
\providecommand \selectlanguage [0]{\@gobble}%
\providecommand \bibinfo  [0]{\@secondoftwo}%
\providecommand \bibfield  [0]{\@secondoftwo}%
\providecommand \translation [1]{[#1]}%
\providecommand \BibitemOpen [0]{}%
\providecommand \bibitemStop [0]{}%
\providecommand \bibitemNoStop [0]{.\EOS\space}%
\providecommand \EOS [0]{\spacefactor3000\relax}%
\providecommand \BibitemShut  [1]{\csname bibitem#1\endcsname}%
\let\auto@bib@innerbib\@empty
%</preamble>
\bibitem [{\citenamefont {Gayathri}\ \emph {et~al.}(2020)\citenamefont
  {Gayathri}, \citenamefont {Lopez}, \citenamefont {R.~S.}, \citenamefont
  {Heng}, \citenamefont {Pai},\ and\ \citenamefont {Messenger}}]{gmm}%
  \BibitemOpen
  \bibfield  {author} {\bibinfo {author} {\bibfnamefont {V.}~\bibnamefont
  {Gayathri}}, \bibinfo {author} {\bibfnamefont {D.}~\bibnamefont {Lopez}},
  \bibinfo {author} {\bibfnamefont {P.}~\bibnamefont {R.~S.}}, \bibinfo
  {author} {\bibfnamefont {I.~S.}\ \bibnamefont {Heng}}, \bibinfo {author}
  {\bibfnamefont {A.}~\bibnamefont {Pai}}, \ and\ \bibinfo {author}
  {\bibfnamefont {C.}~\bibnamefont {Messenger}},\ }\href {\doibase
  10.1103/PhysRevD.102.104023} {\bibfield  {journal} {\bibinfo  {journal}
  {Phys. Rev. D}\ }\textbf {\bibinfo {volume} {102}},\ \bibinfo {pages}
  {104023} (\bibinfo {year} {2020})}\BibitemShut {NoStop}%
\bibitem [{\citenamefont {Abbott}\ \emph
  {et~al.}(2021{\natexlab{a}})\citenamefont {Abbott} \emph {et~al.}}]{gwtc2}%
  \BibitemOpen
  \bibfield  {author} {\bibinfo {author} {\bibfnamefont {B.~P.}\ \bibnamefont
  {Abbott}} \emph {et~al.} (\bibinfo {collaboration} {LIGO Scientific
  Collaboration and Virgo Collaboration}),\ }\href {\doibase
  10.1103/PhysRevX.11.021053} {\bibfield  {journal} {\bibinfo  {journal} {Phys.
  Rev. X}\ }\textbf {\bibinfo {volume} {11}},\ \bibinfo {pages} {021053}
  (\bibinfo {year} {2021}{\natexlab{a}})}\BibitemShut {NoStop}%
\bibitem [{\citenamefont {Abbott}\ \emph
  {et~al.}(2021{\natexlab{b}})\citenamefont {Abbott} \emph {et~al.}}]{gwtc2_1}%
  \BibitemOpen
  \bibfield  {author} {\bibinfo {author} {\bibfnamefont {R.}~\bibnamefont
  {Abbott}} \emph {et~al.} (\bibinfo {collaboration} {LIGO Scientific,
  VIRGO}),\ }\href@noop {} {\  (\bibinfo {year} {2021}{\natexlab{b}})},\
  \Eprint {http://arxiv.org/abs/2108.01045} {arXiv:2108.01045 [gr-qc]}
  \BibitemShut {NoStop}%
\bibitem [{\citenamefont {Aasi}\ \emph {et~al.}(2015)\citenamefont {Aasi} \emph
  {et~al.}}]{LIGO-det}%
  \BibitemOpen
  \bibfield  {author} {\bibinfo {author} {\bibfnamefont {J.}~\bibnamefont
  {Aasi}} \emph {et~al.} (\bibinfo {collaboration} {LIGO Scientific}),\ }\href
  {\doibase 10.1088/0264-9381/32/7/074001} {\bibfield  {journal} {\bibinfo
  {journal} {Class. Quant. Grav.}\ }\textbf {\bibinfo {volume} {32}},\ \bibinfo
  {pages} {074001} (\bibinfo {year} {2015})},\ \Eprint
  {http://arxiv.org/abs/1411.4547} {arXiv:1411.4547 [gr-qc]} \BibitemShut
  {NoStop}%
\bibitem [{\citenamefont {Acernese}\ \emph {et~al.}(2015)\citenamefont
  {Acernese} \emph {et~al.}}]{Virgo-det}%
  \BibitemOpen
  \bibfield  {author} {\bibinfo {author} {\bibfnamefont {F.}~\bibnamefont
  {Acernese}} \emph {et~al.} (\bibinfo {collaboration} {Virgo Collaboration}),\
  }\href {\doibase 10.1088/0264-9381/32/2/024001} {\bibfield  {journal}
  {\bibinfo  {journal} {Class. Quant. Grav.}\ }\textbf {\bibinfo {volume}
  {32}},\ \bibinfo {pages} {024001} (\bibinfo {year} {2015})}\BibitemShut
  {NoStop}%
\bibitem [{\citenamefont {Abbott}\ \emph
  {et~al.}(2020{\natexlab{a}})\citenamefont {Abbott} \emph
  {et~al.}}]{GW190521}%
  \BibitemOpen
  \bibfield  {author} {\bibinfo {author} {\bibfnamefont {R.}~\bibnamefont
  {Abbott}} \emph {et~al.} (\bibinfo {collaboration} {LIGO Scientific
  Collaboration and Virgo Collaboration}),\ }\href {\doibase
  10.1103/PhysRevLett.125.101102} {\bibfield  {journal} {\bibinfo  {journal}
  {Phys. Rev. Lett.}\ }\textbf {\bibinfo {volume} {125}},\ \bibinfo {pages}
  {101102} (\bibinfo {year} {2020}{\natexlab{a}})}\BibitemShut {NoStop}%
\bibitem [{\citenamefont {Abbott}\ \emph
  {et~al.}(2020{\natexlab{b}})\citenamefont {Abbott} \emph
  {et~al.}}]{GW190814}%
  \BibitemOpen
  \bibfield  {author} {\bibinfo {author} {\bibfnamefont {R.}~\bibnamefont
  {Abbott}} \emph {et~al.} (\bibinfo {collaboration} {LIGO Scientific,
  Virgo}),\ }\href {\doibase 10.3847/2041-8213/ab960f} {\bibfield  {journal}
  {\bibinfo  {journal} {Astrophys. J. Lett.}\ }\textbf {\bibinfo {volume}
  {896}},\ \bibinfo {pages} {L44} (\bibinfo {year} {2020}{\natexlab{b}})},\
  \Eprint {http://arxiv.org/abs/2006.12611} {arXiv:2006.12611 [astro-ph.HE]}
  \BibitemShut {NoStop}%
\bibitem [{\citenamefont {Abbott}\ \emph
  {et~al.}(2020{\natexlab{c}})\citenamefont {Abbott} \emph
  {et~al.}}]{GW190425}%
  \BibitemOpen
  \bibfield  {author} {\bibinfo {author} {\bibfnamefont {B.~P.}\ \bibnamefont
  {Abbott}} \emph {et~al.} (\bibinfo {collaboration} {LIGO Scientific,
  Virgo}),\ }\href {\doibase 10.3847/2041-8213/ab75f5} {\bibfield  {journal}
  {\bibinfo  {journal} {Astrophys. J. Lett.}\ }\textbf {\bibinfo {volume}
  {892}},\ \bibinfo {pages} {L3} (\bibinfo {year} {2020}{\natexlab{c}})},\
  \Eprint {http://arxiv.org/abs/2001.01761} {arXiv:2001.01761 [astro-ph.HE]}
  \BibitemShut {NoStop}%
\bibitem [{\citenamefont {Abbott}\ \emph
  {et~al.}(2021{\natexlab{c}})\citenamefont {Abbott} \emph {et~al.}}]{NSBH}%
  \BibitemOpen
  \bibfield  {author} {\bibinfo {author} {\bibfnamefont {R.}~\bibnamefont
  {Abbott}} \emph {et~al.} (\bibinfo {collaboration} {LIGO Scientific, KAGRA,
  VIRGO}),\ }\href {\doibase 10.3847/2041-8213/ac082e} {\bibfield  {journal}
  {\bibinfo  {journal} {Astrophys. J. Lett.}\ }\textbf {\bibinfo {volume}
  {915}},\ \bibinfo {pages} {L5} (\bibinfo {year} {2021}{\natexlab{c}})},\
  \Eprint {http://arxiv.org/abs/2106.15163} {arXiv:2106.15163 [astro-ph.HE]}
  \BibitemShut {NoStop}%
\bibitem [{\citenamefont {Abbott}\ \emph
  {et~al.}(2021{\natexlab{d}})\citenamefont {Abbott} \emph {et~al.}}]{gwtc3}%
  \BibitemOpen
  \bibfield  {author} {\bibinfo {author} {\bibfnamefont {R.}~\bibnamefont
  {Abbott}} \emph {et~al.} (\bibinfo {collaboration} {LIGO Scientific, VIRGO,
  KAGRA}),\ }\href@noop {} {\  (\bibinfo {year} {2021}{\natexlab{d}})},\
  \Eprint {http://arxiv.org/abs/2111.03606} {arXiv:2111.03606 [gr-qc]}
  \BibitemShut {NoStop}%
\bibitem [{\citenamefont {Abbott}\ \emph
  {et~al.}(2020{\natexlab{d}})\citenamefont {Abbott} \emph
  {et~al.}}]{ccsn_ligo}%
  \BibitemOpen
  \bibfield  {author} {\bibinfo {author} {\bibfnamefont {B.~P.}\ \bibnamefont
  {Abbott}} \emph {et~al.} (\bibinfo {collaboration} {LIGO Scientific,
  Virgo}),\ }\href {\doibase 10.1103/PhysRevD.101.084002} {\bibfield  {journal}
  {\bibinfo  {journal} {Phys. Rev. D}\ }\textbf {\bibinfo {volume} {101}},\
  \bibinfo {pages} {084002} (\bibinfo {year} {2020}{\natexlab{d}})},\ \Eprint
  {http://arxiv.org/abs/1908.03584} {arXiv:1908.03584 [astro-ph.HE]}
  \BibitemShut {NoStop}%
\bibitem [{\citenamefont {Abbott}\ \emph
  {et~al.}(2021{\natexlab{e}})\citenamefont {Abbott} \emph
  {et~al.}}]{cosmic_string}%
  \BibitemOpen
  \bibfield  {author} {\bibinfo {author} {\bibfnamefont {R.}~\bibnamefont
  {Abbott}} \emph {et~al.} (\bibinfo {collaboration} {LIGO Scientific, Virgo,
  KAGRA}),\ }\href {\doibase 10.1103/PhysRevLett.126.241102} {\bibfield
  {journal} {\bibinfo  {journal} {Phys. Rev. Lett.}\ }\textbf {\bibinfo
  {volume} {126}},\ \bibinfo {pages} {241102} (\bibinfo {year}
  {2021}{\natexlab{e}})},\ \Eprint {http://arxiv.org/abs/2101.12248}
  {arXiv:2101.12248 [gr-qc]} \BibitemShut {NoStop}%
\bibitem [{\citenamefont {Ebersold}\ and\ \citenamefont
  {Tiwari}(2020)}]{memory}%
  \BibitemOpen
  \bibfield  {author} {\bibinfo {author} {\bibfnamefont {M.}~\bibnamefont
  {Ebersold}}\ and\ \bibinfo {author} {\bibfnamefont {S.}~\bibnamefont
  {Tiwari}},\ }\href {\doibase 10.1103/PhysRevD.101.104041} {\bibfield
  {journal} {\bibinfo  {journal} {Phys. Rev. D}\ }\textbf {\bibinfo {volume}
  {101}},\ \bibinfo {pages} {104041} (\bibinfo {year} {2020})},\ \Eprint
  {http://arxiv.org/abs/2005.03306} {arXiv:2005.03306 [gr-qc]} \BibitemShut
  {NoStop}%
\bibitem [{\citenamefont {Abbott}\ \emph
  {et~al.}(2019{\natexlab{a}})\citenamefont {Abbott} \emph
  {et~al.}}]{magnetar}%
  \BibitemOpen
  \bibfield  {author} {\bibinfo {author} {\bibfnamefont {B.~P.}\ \bibnamefont
  {Abbott}} \emph {et~al.} (\bibinfo {collaboration} {LIGO Scientific,
  Virgo}),\ }\href {\doibase 10.3847/1538-4357/ab0e15} {\bibfield  {journal}
  {\bibinfo  {journal} {Astrophys. J.}\ }\textbf {\bibinfo {volume} {874}},\
  \bibinfo {pages} {163} (\bibinfo {year} {2019}{\natexlab{a}})},\ \Eprint
  {http://arxiv.org/abs/1902.01557} {arXiv:1902.01557 [astro-ph.HE]}
  \BibitemShut {NoStop}%
\bibitem [{\citenamefont {Abadie}\ \emph {et~al.}(2011)\citenamefont {Abadie}
  \emph {et~al.}}]{ns_glitch}%
  \BibitemOpen
  \bibfield  {author} {\bibinfo {author} {\bibfnamefont {J.}~\bibnamefont
  {Abadie}} \emph {et~al.} (\bibinfo {collaboration} {LIGO Scientific}),\
  }\href {\doibase 10.1103/PhysRevD.83.042001} {\bibfield  {journal} {\bibinfo
  {journal} {Phys. Rev. D}\ }\textbf {\bibinfo {volume} {83}},\ \bibinfo
  {pages} {042001} (\bibinfo {year} {2011})},\ \Eprint
  {http://arxiv.org/abs/1011.1357} {arXiv:1011.1357 [gr-qc]} \BibitemShut
  {NoStop}%
\bibitem [{\citenamefont {Abadie}\ \emph {et~al.}(2010)\citenamefont {Abadie}
  \emph {et~al.}}]{allskyfith}%
  \BibitemOpen
  \bibfield  {author} {\bibinfo {author} {\bibfnamefont {J.}~\bibnamefont
  {Abadie}} \emph {et~al.} (\bibinfo {collaboration} {The LIGO Scientific
  Collaboration and The Virgo Collaboration}),\ }\href {\doibase
  10.1103/PhysRevD.81.102001} {\bibfield  {journal} {\bibinfo  {journal} {Phys.
  Rev. D}\ }\textbf {\bibinfo {volume} {81}},\ \bibinfo {pages} {102001}
  (\bibinfo {year} {2010})}\BibitemShut {NoStop}%
\bibitem [{\citenamefont {Abadie}\ \emph {et~al.}(2012)\citenamefont {Abadie}
  \emph {et~al.}}]{allskysixth}%
  \BibitemOpen
  \bibfield  {author} {\bibinfo {author} {\bibfnamefont {J.}~\bibnamefont
  {Abadie}} \emph {et~al.} (\bibinfo {collaboration} {The LIGO Scientific
  Collaboration and The Virgo Collaboration}),\ }\href {\doibase
  10.1103/PhysRevD.85.122007} {\bibfield  {journal} {\bibinfo  {journal} {Phys.
  Rev. D}\ }\textbf {\bibinfo {volume} {85}},\ \bibinfo {pages} {122007}
  (\bibinfo {year} {2012})}\BibitemShut {NoStop}%
\bibitem [{\citenamefont {Abbott}\ \emph {et~al.}(2017)\citenamefont {Abbott}
  \emph {et~al.}}]{allakyo1}%
  \BibitemOpen
  \bibfield  {author} {\bibinfo {author} {\bibfnamefont {B.~P.}\ \bibnamefont
  {Abbott}} \emph {et~al.} (\bibinfo {collaboration} {LIGO Scientific
  Collaboration and Virgo Collaboration}),\ }\href {\doibase
  10.1103/PhysRevD.95.042003} {\bibfield  {journal} {\bibinfo  {journal} {Phys.
  Rev. D}\ }\textbf {\bibinfo {volume} {95}},\ \bibinfo {pages} {042003}
  (\bibinfo {year} {2017})}\BibitemShut {NoStop}%
\bibitem [{\citenamefont {Abbott}\ \emph
  {et~al.}(2019{\natexlab{b}})\citenamefont {Abbott} \emph {et~al.}}]{alskyo2}%
  \BibitemOpen
  \bibfield  {author} {\bibinfo {author} {\bibfnamefont {B.~P.}\ \bibnamefont
  {Abbott}} \emph {et~al.} (\bibinfo {collaboration} {LIGO Scientific
  Collaboration and Virgo Collaboration}),\ }\href {\doibase
  10.1103/PhysRevD.100.024017} {\bibfield  {journal} {\bibinfo  {journal}
  {Phys. Rev. D}\ }\textbf {\bibinfo {volume} {100}},\ \bibinfo {pages}
  {024017} (\bibinfo {year} {2019}{\natexlab{b}})}\BibitemShut {NoStop}%
\bibitem [{\citenamefont {Abbott}\ \emph
  {et~al.}(2021{\natexlab{f}})\citenamefont {Abbott} \emph
  {et~al.}}]{allskyo3}%
  \BibitemOpen
  \bibfield  {author} {\bibinfo {author} {\bibfnamefont {R.}~\bibnamefont
  {Abbott}} \emph {et~al.} (\bibinfo {collaboration} {KAGRA, VIRGO, LIGO
  Scientific}),\ }\href {\doibase 10.1103/PhysRevD.104.122004} {\bibfield
  {journal} {\bibinfo  {journal} {Phys. Rev. D}\ }\textbf {\bibinfo {volume}
  {104}},\ \bibinfo {pages} {122004} (\bibinfo {year} {2021}{\natexlab{f}})},\
  \Eprint {http://arxiv.org/abs/2107.03701} {arXiv:2107.03701 [gr-qc]}
  \BibitemShut {NoStop}%
\bibitem [{\citenamefont {Klimenko}\ \emph {et~al.}(2008)\citenamefont
  {Klimenko}, \citenamefont {Yakushin}, \citenamefont {Mercer},\ and\
  \citenamefont {Mitselmakher}}]{cwb_2008}%
  \BibitemOpen
  \bibfield  {author} {\bibinfo {author} {\bibfnamefont {S.}~\bibnamefont
  {Klimenko}}, \bibinfo {author} {\bibfnamefont {I.}~\bibnamefont {Yakushin}},
  \bibinfo {author} {\bibfnamefont {A.}~\bibnamefont {Mercer}}, \ and\ \bibinfo
  {author} {\bibfnamefont {G.}~\bibnamefont {Mitselmakher}},\ }\href {\doibase
  10.1088/0264-9381/25/11/114029} {\bibfield  {journal} {\bibinfo  {journal}
  {Class. Quant. Grav.}\ }\textbf {\bibinfo {volume} {25}},\ \bibinfo {pages}
  {114029} (\bibinfo {year} {2008})},\ \Eprint {http://arxiv.org/abs/0802.3232}
  {arXiv:0802.3232 [gr-qc]} \BibitemShut {NoStop}%
\bibitem [{\citenamefont {Abbott}\ \emph {et~al.}(2016)\citenamefont {Abbott}
  \emph {et~al.}}]{gw150914_cwb}%
  \BibitemOpen
  \bibfield  {author} {\bibinfo {author} {\bibfnamefont {B.~P.}\ \bibnamefont
  {Abbott}} \emph {et~al.} (\bibinfo {collaboration} {LIGO Scientific
  Collaboration and Virgo Collaboration}),\ }\href {\doibase
  10.1103/PhysRevD.93.122004} {\bibfield  {journal} {\bibinfo  {journal} {Phys.
  Rev. D}\ }\textbf {\bibinfo {volume} {93}},\ \bibinfo {pages} {122004}
  (\bibinfo {year} {2016})}\BibitemShut {NoStop}%
\bibitem [{\citenamefont {Szczepa\ifmmode~\acute{n}\else \'{n}\fi{}czyk}\ \emph
  {et~al.}(2021)\citenamefont {Szczepa\ifmmode~\acute{n}\else \'{n}\fi{}czyk},
  \citenamefont {Klimenko}, \citenamefont {O'Brien}, \citenamefont {Bartos},
  \citenamefont {Gayathri}, \citenamefont {Mitselmakher}, \citenamefont
  {Prodi}, \citenamefont {Vedovato}, \citenamefont {Lazzaro}, \citenamefont
  {Milotti}, \citenamefont {Salemi}, \citenamefont {Drago},\ and\ \citenamefont
  {Tiwari}}]{gw190425_cwb}%
  \BibitemOpen
  \bibfield  {author} {\bibinfo {author} {\bibfnamefont {M.}~\bibnamefont
  {Szczepa\ifmmode~\acute{n}\else \'{n}\fi{}czyk}}, \bibinfo {author}
  {\bibfnamefont {S.}~\bibnamefont {Klimenko}}, \bibinfo {author}
  {\bibfnamefont {B.}~\bibnamefont {O'Brien}}, \bibinfo {author} {\bibfnamefont
  {I.}~\bibnamefont {Bartos}}, \bibinfo {author} {\bibfnamefont
  {V.}~\bibnamefont {Gayathri}}, \bibinfo {author} {\bibfnamefont
  {G.}~\bibnamefont {Mitselmakher}}, \bibinfo {author} {\bibfnamefont
  {G.}~\bibnamefont {Prodi}}, \bibinfo {author} {\bibfnamefont
  {G.}~\bibnamefont {Vedovato}}, \bibinfo {author} {\bibfnamefont
  {C.}~\bibnamefont {Lazzaro}}, \bibinfo {author} {\bibfnamefont
  {E.}~\bibnamefont {Milotti}}, \bibinfo {author} {\bibfnamefont
  {F.}~\bibnamefont {Salemi}}, \bibinfo {author} {\bibfnamefont
  {M.}~\bibnamefont {Drago}}, \ and\ \bibinfo {author} {\bibfnamefont
  {S.}~\bibnamefont {Tiwari}},\ }\href {\doibase 10.1103/PhysRevD.103.082002}
  {\bibfield  {journal} {\bibinfo  {journal} {Phys. Rev. D}\ }\textbf {\bibinfo
  {volume} {103}},\ \bibinfo {pages} {082002} (\bibinfo {year}
  {2021})}\BibitemShut {NoStop}%
\bibitem [{\citenamefont {Powell}\ \emph {et~al.}(2015)\citenamefont {Powell},
  \citenamefont {Trifir\`o}, \citenamefont {Cuoco}, \citenamefont {Heng},\ and\
  \citenamefont {Cavagli\`a}}]{noise_cl1}%
  \BibitemOpen
  \bibfield  {author} {\bibinfo {author} {\bibfnamefont {J.}~\bibnamefont
  {Powell}}, \bibinfo {author} {\bibfnamefont {D.}~\bibnamefont {Trifir\`o}},
  \bibinfo {author} {\bibfnamefont {E.}~\bibnamefont {Cuoco}}, \bibinfo
  {author} {\bibfnamefont {I.~S.}\ \bibnamefont {Heng}}, \ and\ \bibinfo
  {author} {\bibfnamefont {M.}~\bibnamefont {Cavagli\`a}},\ }\href {\doibase
  10.1088/0264-9381/32/21/215012} {\bibfield  {journal} {\bibinfo  {journal}
  {Class. Quant. Grav.}\ }\textbf {\bibinfo {volume} {32}},\ \bibinfo {pages}
  {215012} (\bibinfo {year} {2015})},\ \Eprint
  {http://arxiv.org/abs/1505.01299} {arXiv:1505.01299 [astro-ph.IM]}
  \BibitemShut {NoStop}%
\bibitem [{\citenamefont {Powell}\ \emph {et~al.}(2017)\citenamefont {Powell},
  \citenamefont {Torres-Forn\'e}, \citenamefont {Lynch}, \citenamefont
  {Trifir\`o}, \citenamefont {Cuoco}, \citenamefont {Cavagli\`a}, \citenamefont
  {Heng},\ and\ \citenamefont {Font}}]{noise_cl2}%
  \BibitemOpen
  \bibfield  {author} {\bibinfo {author} {\bibfnamefont {J.}~\bibnamefont
  {Powell}}, \bibinfo {author} {\bibfnamefont {A.}~\bibnamefont
  {Torres-Forn\'e}}, \bibinfo {author} {\bibfnamefont {R.}~\bibnamefont
  {Lynch}}, \bibinfo {author} {\bibfnamefont {D.}~\bibnamefont {Trifir\`o}},
  \bibinfo {author} {\bibfnamefont {E.}~\bibnamefont {Cuoco}}, \bibinfo
  {author} {\bibfnamefont {M.}~\bibnamefont {Cavagli\`a}}, \bibinfo {author}
  {\bibfnamefont {I.~S.}\ \bibnamefont {Heng}}, \ and\ \bibinfo {author}
  {\bibfnamefont {J.~A.}\ \bibnamefont {Font}},\ }\href {\doibase
  10.1088/1361-6382/34/3/034002} {\bibfield  {journal} {\bibinfo  {journal}
  {Class. Quant. Grav.}\ }\textbf {\bibinfo {volume} {34}},\ \bibinfo {pages}
  {034002} (\bibinfo {year} {2017})},\ \Eprint
  {http://arxiv.org/abs/1609.06262} {arXiv:1609.06262 [astro-ph.IM]}
  \BibitemShut {NoStop}%
\bibitem [{\citenamefont {Mukund}\ \emph {et~al.}(2017)\citenamefont {Mukund},
  \citenamefont {Abraham}, \citenamefont {Kandhasamy}, \citenamefont {Mitra},\
  and\ \citenamefont {Philip}}]{noise_cl3}%
  \BibitemOpen
  \bibfield  {author} {\bibinfo {author} {\bibfnamefont {N.}~\bibnamefont
  {Mukund}}, \bibinfo {author} {\bibfnamefont {S.}~\bibnamefont {Abraham}},
  \bibinfo {author} {\bibfnamefont {S.}~\bibnamefont {Kandhasamy}}, \bibinfo
  {author} {\bibfnamefont {S.}~\bibnamefont {Mitra}}, \ and\ \bibinfo {author}
  {\bibfnamefont {N.~S.}\ \bibnamefont {Philip}},\ }\href {\doibase
  10.1103/PhysRevD.95.104059} {\bibfield  {journal} {\bibinfo  {journal} {Phys.
  Rev. D}\ }\textbf {\bibinfo {volume} {95}},\ \bibinfo {pages} {104059}
  (\bibinfo {year} {2017})}\BibitemShut {NoStop}%
\bibitem [{\citenamefont {George}\ \emph {et~al.}(2018)\citenamefont {George},
  \citenamefont {Shen},\ and\ \citenamefont {Huerta}}]{noise_cl4}%
  \BibitemOpen
  \bibfield  {author} {\bibinfo {author} {\bibfnamefont {D.}~\bibnamefont
  {George}}, \bibinfo {author} {\bibfnamefont {H.}~\bibnamefont {Shen}}, \ and\
  \bibinfo {author} {\bibfnamefont {E.~A.}\ \bibnamefont {Huerta}},\ }\href
  {\doibase 10.1103/PhysRevD.97.101501} {\bibfield  {journal} {\bibinfo
  {journal} {Phys. Rev. D}\ }\textbf {\bibinfo {volume} {97}},\ \bibinfo
  {pages} {101501} (\bibinfo {year} {2018})}\BibitemShut {NoStop}%
\bibitem [{\citenamefont {Biswas}\ \emph {et~al.}(2013)\citenamefont {Biswas}
  \emph {et~al.}}]{noise_cl5}%
  \BibitemOpen
  \bibfield  {author} {\bibinfo {author} {\bibfnamefont {R.}~\bibnamefont
  {Biswas}} \emph {et~al.},\ }\href {\doibase 10.1103/PhysRevD.88.062003}
  {\bibfield  {journal} {\bibinfo  {journal} {Phys. Rev. D}\ }\textbf {\bibinfo
  {volume} {88}},\ \bibinfo {pages} {062003} (\bibinfo {year}
  {2013})}\BibitemShut {NoStop}%
\bibitem [{\citenamefont {Rampone}\ \emph {et~al.}(2013)\citenamefont
  {Rampone}, \citenamefont {Pierro}, \citenamefont {Troiano},\ and\
  \citenamefont {Pinto}}]{noise_cl6}%
  \BibitemOpen
  \bibfield  {author} {\bibinfo {author} {\bibfnamefont {S.}~\bibnamefont
  {Rampone}}, \bibinfo {author} {\bibfnamefont {V.}~\bibnamefont {Pierro}},
  \bibinfo {author} {\bibfnamefont {L.}~\bibnamefont {Troiano}}, \ and\
  \bibinfo {author} {\bibfnamefont {I.~M.}\ \bibnamefont {Pinto}},\ }\href
  {\doibase 10.1142/S0129183113500848} {\bibfield  {journal} {\bibinfo
  {journal} {Int. J. Mod. Phys. C}\ }\textbf {\bibinfo {volume} {24}},\
  \bibinfo {pages} {1350084} (\bibinfo {year} {2013})},\ \Eprint
  {http://arxiv.org/abs/1401.5941} {arXiv:1401.5941 [astro-ph.IM]} \BibitemShut
  {NoStop}%
\bibitem [{\citenamefont {Lightman}\ \emph {et~al.}(2006)\citenamefont
  {Lightman}, \citenamefont {Thurakal}, \citenamefont {Dwyer}, \citenamefont
  {Grossman}, \citenamefont {Kalmus}, \citenamefont {Matone}, \citenamefont
  {Rollins}, \citenamefont {Zairis},\ and\ \citenamefont {Marka}}]{noise_cl7}%
  \BibitemOpen
  \bibfield  {author} {\bibinfo {author} {\bibfnamefont {M.}~\bibnamefont
  {Lightman}}, \bibinfo {author} {\bibfnamefont {J.}~\bibnamefont {Thurakal}},
  \bibinfo {author} {\bibfnamefont {J.}~\bibnamefont {Dwyer}}, \bibinfo
  {author} {\bibfnamefont {R.}~\bibnamefont {Grossman}}, \bibinfo {author}
  {\bibfnamefont {P.}~\bibnamefont {Kalmus}}, \bibinfo {author} {\bibfnamefont
  {L.}~\bibnamefont {Matone}}, \bibinfo {author} {\bibfnamefont
  {J.}~\bibnamefont {Rollins}}, \bibinfo {author} {\bibfnamefont
  {S.}~\bibnamefont {Zairis}}, \ and\ \bibinfo {author} {\bibfnamefont
  {S.}~\bibnamefont {Marka}},\ }\href {\doibase 10.1088/1742-6596/32/1/010}
  {\bibfield  {journal} {\bibinfo  {journal} {J. Phys. Conf. Ser.}\ }\textbf
  {\bibinfo {volume} {32}},\ \bibinfo {pages} {58} (\bibinfo {year}
  {2006})}\BibitemShut {NoStop}%
\bibitem [{\citenamefont {Razzano}\ and\ \citenamefont
  {Cuoco}(2018)}]{noise_cl8}%
  \BibitemOpen
  \bibfield  {author} {\bibinfo {author} {\bibfnamefont {M.}~\bibnamefont
  {Razzano}}\ and\ \bibinfo {author} {\bibfnamefont {E.}~\bibnamefont
  {Cuoco}},\ }\href {\doibase 10.1088/1361-6382/aab793} {\bibfield  {journal}
  {\bibinfo  {journal} {Class. Quant. Grav.}\ }\textbf {\bibinfo {volume}
  {35}},\ \bibinfo {pages} {095016} (\bibinfo {year} {2018})},\ \Eprint
  {http://arxiv.org/abs/1803.09933} {arXiv:1803.09933 [gr-qc]} \BibitemShut
  {NoStop}%
\bibitem [{\citenamefont {Huerta}\ \emph {et~al.}(2021)\citenamefont {Huerta}
  \emph {et~al.}}]{noise_cl9}%
  \BibitemOpen
  \bibfield  {author} {\bibinfo {author} {\bibfnamefont {E.~A.}\ \bibnamefont
  {Huerta}} \emph {et~al.},\ }\href {\doibase 10.1038/s41550-021-01405-0}
  {\bibfield  {journal} {\bibinfo  {journal} {Nature Astron.}\ }\textbf
  {\bibinfo {volume} {5}},\ \bibinfo {pages} {1062} (\bibinfo {year} {2021})},\
  \Eprint {http://arxiv.org/abs/2012.08545} {arXiv:2012.08545 [gr-qc]}
  \BibitemShut {NoStop}%
\bibitem [{\citenamefont {Cavaglia}\ \emph
  {et~al.}(2020{\natexlab{a}})\citenamefont {Cavaglia}, \citenamefont {Gaudio},
  \citenamefont {Hansen}, \citenamefont {Staats}, \citenamefont
  {Szczepanczyk},\ and\ \citenamefont {Zanolin}}]{noise_cl10}%
  \BibitemOpen
  \bibfield  {author} {\bibinfo {author} {\bibfnamefont {M.}~\bibnamefont
  {Cavaglia}}, \bibinfo {author} {\bibfnamefont {S.}~\bibnamefont {Gaudio}},
  \bibinfo {author} {\bibfnamefont {T.}~\bibnamefont {Hansen}}, \bibinfo
  {author} {\bibfnamefont {K.}~\bibnamefont {Staats}}, \bibinfo {author}
  {\bibfnamefont {M.}~\bibnamefont {Szczepanczyk}}, \ and\ \bibinfo {author}
  {\bibfnamefont {M.}~\bibnamefont {Zanolin}},\ }\href {\doibase
  10.1088/2632-2153/ab527d} {\bibfield  {journal} {\bibinfo  {journal} {Mach.
  Learn. Sci. Tech.}\ }\textbf {\bibinfo {volume} {1}},\ \bibinfo {pages}
  {015005} (\bibinfo {year} {2020}{\natexlab{a}})},\ \Eprint
  {http://arxiv.org/abs/2002.04591} {arXiv:2002.04591 [astro-ph.IM]}
  \BibitemShut {NoStop}%
\bibitem [{\citenamefont {Vinciguerra}\ \emph {et~al.}(2017)\citenamefont
  {Vinciguerra}, \citenamefont {Drago}, \citenamefont {Prodi}, \citenamefont
  {Klimenko}, \citenamefont {Lazzaro}, \citenamefont {Necula}, \citenamefont
  {Salemi}, \citenamefont {Tiwari}, \citenamefont {Tringali},\ and\
  \citenamefont {Vedovato}}]{cwb_ml1}%
  \BibitemOpen
  \bibfield  {author} {\bibinfo {author} {\bibfnamefont {S.}~\bibnamefont
  {Vinciguerra}}, \bibinfo {author} {\bibfnamefont {M.}~\bibnamefont {Drago}},
  \bibinfo {author} {\bibfnamefont {G.~A.}\ \bibnamefont {Prodi}}, \bibinfo
  {author} {\bibfnamefont {S.}~\bibnamefont {Klimenko}}, \bibinfo {author}
  {\bibfnamefont {C.}~\bibnamefont {Lazzaro}}, \bibinfo {author} {\bibfnamefont
  {V.}~\bibnamefont {Necula}}, \bibinfo {author} {\bibfnamefont
  {F.}~\bibnamefont {Salemi}}, \bibinfo {author} {\bibfnamefont
  {V.}~\bibnamefont {Tiwari}}, \bibinfo {author} {\bibfnamefont {M.~C.}\
  \bibnamefont {Tringali}}, \ and\ \bibinfo {author} {\bibfnamefont
  {G.}~\bibnamefont {Vedovato}},\ }\href {\doibase 10.1088/1361-6382/aa6654}
  {\bibfield  {journal} {\bibinfo  {journal} {Class. Quant. Grav.}\ }\textbf
  {\bibinfo {volume} {34}},\ \bibinfo {pages} {094003} (\bibinfo {year}
  {2017})},\ \Eprint {http://arxiv.org/abs/1702.03208} {arXiv:1702.03208
  [astro-ph.IM]} \BibitemShut {NoStop}%
\bibitem [{\citenamefont {Cavaglia}\ \emph
  {et~al.}(2020{\natexlab{b}})\citenamefont {Cavaglia}, \citenamefont {Gaudio},
  \citenamefont {Hansen}, \citenamefont {Staats}, \citenamefont
  {Szczepanczyk},\ and\ \citenamefont {Zanolin}}]{cwb_ml2}%
  \BibitemOpen
  \bibfield  {author} {\bibinfo {author} {\bibfnamefont {M.}~\bibnamefont
  {Cavaglia}}, \bibinfo {author} {\bibfnamefont {S.}~\bibnamefont {Gaudio}},
  \bibinfo {author} {\bibfnamefont {T.}~\bibnamefont {Hansen}}, \bibinfo
  {author} {\bibfnamefont {K.}~\bibnamefont {Staats}}, \bibinfo {author}
  {\bibfnamefont {M.}~\bibnamefont {Szczepanczyk}}, \ and\ \bibinfo {author}
  {\bibfnamefont {M.}~\bibnamefont {Zanolin}},\ }\href {\doibase
  10.1088/2632-2153/ab527d} {\bibfield  {journal} {\bibinfo  {journal} {Mach.
  Learn. Sci. Tech.}\ }\textbf {\bibinfo {volume} {1}},\ \bibinfo {pages}
  {015005} (\bibinfo {year} {2020}{\natexlab{b}})},\ \Eprint
  {http://arxiv.org/abs/2002.04591} {arXiv:2002.04591 [astro-ph.IM]}
  \BibitemShut {NoStop}%
\bibitem [{\citenamefont {Mishra}\ \emph {et~al.}(2021)\citenamefont {Mishra},
  \citenamefont {O'Brien}, \citenamefont {Gayathri}, \citenamefont
  {Szczepanczyk}, \citenamefont {Bhaumik}, \citenamefont {Bartos},\ and\
  \citenamefont {Klimenko}}]{Mishra2021}%
  \BibitemOpen
  \bibfield  {author} {\bibinfo {author} {\bibfnamefont {T.}~\bibnamefont
  {Mishra}}, \bibinfo {author} {\bibfnamefont {B.}~\bibnamefont {O'Brien}},
  \bibinfo {author} {\bibfnamefont {V.}~\bibnamefont {Gayathri}}, \bibinfo
  {author} {\bibfnamefont {M.}~\bibnamefont {Szczepanczyk}}, \bibinfo {author}
  {\bibfnamefont {S.}~\bibnamefont {Bhaumik}}, \bibinfo {author} {\bibfnamefont
  {I.}~\bibnamefont {Bartos}}, \ and\ \bibinfo {author} {\bibfnamefont
  {S.}~\bibnamefont {Klimenko}},\ }\href {\doibase 10.1103/PhysRevD.104.023014}
  {\bibfield  {journal} {\bibinfo  {journal} {Phys. Rev. D}\ }\textbf {\bibinfo
  {volume} {104}},\ \bibinfo {pages} {023014} (\bibinfo {year} {2021})},\
  \Eprint {http://arxiv.org/abs/2105.04739} {arXiv:2105.04739 [gr-qc]}
  \BibitemShut {NoStop}%
\bibitem [{\citenamefont {Cuoco}\ \emph {et~al.}(2021)\citenamefont {Cuoco}
  \emph {et~al.}}]{Cuoco_2020}%
  \BibitemOpen
  \bibfield  {author} {\bibinfo {author} {\bibfnamefont {E.}~\bibnamefont
  {Cuoco}} \emph {et~al.},\ }\href {\doibase 10.1088/2632-2153/abb93a}
  {\bibfield  {journal} {\bibinfo  {journal} {Mach. Learn. Sci. Tech.}\
  }\textbf {\bibinfo {volume} {2}},\ \bibinfo {pages} {011002} (\bibinfo {year}
  {2021})},\ \Eprint {http://arxiv.org/abs/2005.03745} {arXiv:2005.03745
  [astro-ph.HE]} \BibitemShut {NoStop}%
\bibitem [{\citenamefont {Iess}\ \emph {et~al.}(2020)\citenamefont {Iess},
  \citenamefont {Cuoco}, \citenamefont {Morawski},\ and\ \citenamefont
  {Powell}}]{ml_ccsn}%
  \BibitemOpen
  \bibfield  {author} {\bibinfo {author} {\bibfnamefont {A.}~\bibnamefont
  {Iess}}, \bibinfo {author} {\bibfnamefont {E.}~\bibnamefont {Cuoco}},
  \bibinfo {author} {\bibfnamefont {F.}~\bibnamefont {Morawski}}, \ and\
  \bibinfo {author} {\bibfnamefont {J.}~\bibnamefont {Powell}},\ }\href@noop {}
  {\  (\bibinfo {year} {2020})},\ \Eprint {http://arxiv.org/abs/2001.00279}
  {arXiv:2001.00279 [gr-qc]} \BibitemShut {NoStop}%
\bibitem [{\citenamefont {Necula}\ \emph {et~al.}(2012)\citenamefont {Necula},
  \citenamefont {Klimenko},\ and\ \citenamefont {Mitselmakher}}]{wdm}%
  \BibitemOpen
  \bibfield  {author} {\bibinfo {author} {\bibfnamefont {V.}~\bibnamefont
  {Necula}}, \bibinfo {author} {\bibfnamefont {S.}~\bibnamefont {Klimenko}}, \
  and\ \bibinfo {author} {\bibfnamefont {G.}~\bibnamefont {Mitselmakher}},\
  }\href {\doibase 10.1088/1742-6596/363/1/012032} {\bibfield  {journal}
  {\bibinfo  {journal} {J. Phys. Conf. Ser.}\ }\textbf {\bibinfo {volume}
  {363}},\ \bibinfo {pages} {012032} (\bibinfo {year} {2012})}\BibitemShut
  {NoStop}%
\bibitem [{\citenamefont {Klimenko}\ \emph {et~al.}(2016)\citenamefont
  {Klimenko}, \citenamefont {Vedovato}, \citenamefont {Drago}, \citenamefont
  {Salemi}, \citenamefont {Tiwari}, \citenamefont {Prodi}, \citenamefont
  {Lazzaro}, \citenamefont {Ackley}, \citenamefont {Tiwari}, \citenamefont
  {Da~Silva},\ and\ \citenamefont {Mitselmakher}}]{cwb_Klimenko}%
  \BibitemOpen
  \bibfield  {author} {\bibinfo {author} {\bibfnamefont {S.}~\bibnamefont
  {Klimenko}}, \bibinfo {author} {\bibfnamefont {G.}~\bibnamefont {Vedovato}},
  \bibinfo {author} {\bibfnamefont {M.}~\bibnamefont {Drago}}, \bibinfo
  {author} {\bibfnamefont {F.}~\bibnamefont {Salemi}}, \bibinfo {author}
  {\bibfnamefont {V.}~\bibnamefont {Tiwari}}, \bibinfo {author} {\bibfnamefont
  {G.~A.}\ \bibnamefont {Prodi}}, \bibinfo {author} {\bibfnamefont
  {C.}~\bibnamefont {Lazzaro}}, \bibinfo {author} {\bibfnamefont
  {K.}~\bibnamefont {Ackley}}, \bibinfo {author} {\bibfnamefont
  {S.}~\bibnamefont {Tiwari}}, \bibinfo {author} {\bibfnamefont {C.~F.}\
  \bibnamefont {Da~Silva}}, \ and\ \bibinfo {author} {\bibfnamefont
  {G.}~\bibnamefont {Mitselmakher}},\ }\href {\doibase
  10.1103/PhysRevD.93.042004} {\bibfield  {journal} {\bibinfo  {journal} {Phys.
  Rev. D}\ }\textbf {\bibinfo {volume} {93}},\ \bibinfo {pages} {042004}
  (\bibinfo {year} {2016})}\BibitemShut {NoStop}%
\bibitem [{\citenamefont {\textit{et al}.}(2021)}]{cwb_2021}%
  \BibitemOpen
  \bibfield  {author} {\bibinfo {author} {\bibfnamefont {M.~D.}\ \bibnamefont
  {\textit{et al}.}},\ }\href {\doibase
  https://doi.org/10.1016/j.softx.2021.100678} {\bibfield  {journal} {\bibinfo
  {journal} {SoftwareX}\ }\textbf {\bibinfo {volume} {14}},\ \bibinfo {pages}
  {100678} (\bibinfo {year} {2021})}\BibitemShut {NoStop}%
\bibitem [{cwb()}]{cwb_page}%
  \BibitemOpen
  \href@noop {} {}\bibinfo {howpublished}
  {\url{https://gwburst.gitlab.io}}\BibitemShut {NoStop}%
\bibitem [{\citenamefont {Bishop}(2006)}]{bishop:2006}%
  \BibitemOpen
  \bibfield  {author} {\bibinfo {author} {\bibfnamefont {C.~M.}\ \bibnamefont
  {Bishop}},\ }\href@noop {} {\emph {\bibinfo {title} {Pattern Recognition and
  Machine Learning}}}\ (\bibinfo  {publisher} {Springer},\ \bibinfo {year}
  {2006})\BibitemShut {NoStop}%
\bibitem [{\citenamefont {Schwarz}(1978)}]{bic_78}%
  \BibitemOpen
  \bibfield  {author} {\bibinfo {author} {\bibfnamefont {G.}~\bibnamefont
  {Schwarz}},\ }\href {\doibase 10.1214/aos/1176344136} {\bibfield  {journal}
  {\bibinfo  {journal} {The Annals of Statistics}\ }\textbf {\bibinfo {volume}
  {6}},\ \bibinfo {pages} {461 } (\bibinfo {year} {1978})}\BibitemShut
  {NoStop}%
\bibitem [{\citenamefont {Bhat}\ and\ \citenamefont {Kumar}(2010)}]{bic}%
  \BibitemOpen
  \bibfield  {author} {\bibinfo {author} {\bibfnamefont {H.}~\bibnamefont
  {Bhat}}\ and\ \bibinfo {author} {\bibfnamefont {N.}~\bibnamefont {Kumar}},\
  }\href@noop {} {\  (\bibinfo {year} {2010})}\BibitemShut {NoStop}%
\bibitem [{\citenamefont {{LIGO Scientific Collaboration And Virgo
  Collaboration}}(2021)}]{GWOSC_O3a}%
  \BibitemOpen
  \bibfield  {author} {\bibinfo {author} {\bibnamefont {{LIGO Scientific
  Collaboration And Virgo Collaboration}}},\ }\href {\doibase
  10.7935/nfnt-hm34} {\enquote {\bibinfo {title} {{LIGO} {V}irgo strain data
  from observing run {O}3a},}\ }\bibinfo {howpublished}
  {\url{https://www.gw-openscience.org/O3/O3a}} (\bibinfo {year}
  {2021})\BibitemShut {NoStop}%
\bibitem [{\citenamefont {Abbott}\ \emph
  {et~al.}(2021{\natexlab{g}})\citenamefont {Abbott} \emph {et~al.}}]{GWOSC}%
  \BibitemOpen
  \bibfield  {author} {\bibinfo {author} {\bibfnamefont {R.}~\bibnamefont
  {Abbott}} \emph {et~al.} (\bibinfo {collaboration} {LIGO Scientific
  Collaboration and Virgo Collaboration}),\ }\href {\doibase
  10.1016/j.softx.2021.100658} {\bibfield  {journal} {\bibinfo  {journal}
  {SoftwareX}\ }\textbf {\bibinfo {volume} {13}},\ \bibinfo {pages} {100658}
  (\bibinfo {year} {2021}{\natexlab{g}})}\BibitemShut {NoStop}%
\bibitem [{\citenamefont {Cabero}\ \emph {et~al.}(2019)\citenamefont {Cabero}
  \emph {et~al.}}]{blip_glitch}%
  \BibitemOpen
  \bibfield  {author} {\bibinfo {author} {\bibfnamefont {M.}~\bibnamefont
  {Cabero}} \emph {et~al.},\ }\href {\doibase 10.1088/1361-6382/ab2e14}
  {\bibfield  {journal} {\bibinfo  {journal} {Class. Quant. Grav.}\ }\textbf
  {\bibinfo {volume} {36}},\ \bibinfo {pages} {15} (\bibinfo {year} {2019})},\
  \Eprint {http://arxiv.org/abs/1901.05093} {arXiv:1901.05093
  [physics.ins-det]} \BibitemShut {NoStop}%
\bibitem [{\citenamefont {Sutton}(2013)}]{Sutton_wnb}%
  \BibitemOpen
  \bibfield  {author} {\bibinfo {author} {\bibfnamefont {P.~J.}\ \bibnamefont
  {Sutton}},\ }\href@noop {} {\  (\bibinfo {year} {2013})},\ \Eprint
  {http://arxiv.org/abs/1304.0210} {arXiv:1304.0210 [gr-qc]} \BibitemShut
  {NoStop}%
\bibitem [{\citenamefont {Powell}\ and\ \citenamefont {M\"uller}(2019)}]{s18}%
  \BibitemOpen
  \bibfield  {author} {\bibinfo {author} {\bibfnamefont {J.}~\bibnamefont
  {Powell}}\ and\ \bibinfo {author} {\bibfnamefont {B.}~\bibnamefont
  {M\"uller}},\ }\href {\doibase 10.1093/mnras/stz1304} {\bibfield  {journal}
  {\bibinfo  {journal} {Mon. Not. Roy. Astron. Soc.}\ }\textbf {\bibinfo
  {volume} {487}},\ \bibinfo {pages} {1178} (\bibinfo {year} {2019})},\ \Eprint
  {http://arxiv.org/abs/1812.05738} {arXiv:1812.05738 [astro-ph.HE]}
  \BibitemShut {NoStop}%
\bibitem [{\citenamefont {O'Connor}\ and\ \citenamefont {Couch}(2018)}]{m20}%
  \BibitemOpen
  \bibfield  {author} {\bibinfo {author} {\bibfnamefont {E.~P.}\ \bibnamefont
  {O'Connor}}\ and\ \bibinfo {author} {\bibfnamefont {S.~M.}\ \bibnamefont
  {Couch}},\ }\href {\doibase 10.3847/1538-4357/aadcf7} {\bibfield  {journal}
  {\bibinfo  {journal} {Astrophys. J.}\ }\textbf {\bibinfo {volume} {865}},\
  \bibinfo {pages} {81} (\bibinfo {year} {2018})},\ \Eprint
  {http://arxiv.org/abs/1807.07579} {arXiv:1807.07579 [astro-ph.HE]}
  \BibitemShut {NoStop}%
\bibitem [{\citenamefont {Radice}\ \emph {et~al.}(2019)\citenamefont {Radice},
  \citenamefont {Morozova}, \citenamefont {Burrows}, \citenamefont
  {Vartanyan},\ and\ \citenamefont {Nagakura}}]{s9}%
  \BibitemOpen
  \bibfield  {author} {\bibinfo {author} {\bibfnamefont {D.}~\bibnamefont
  {Radice}}, \bibinfo {author} {\bibfnamefont {V.}~\bibnamefont {Morozova}},
  \bibinfo {author} {\bibfnamefont {A.}~\bibnamefont {Burrows}}, \bibinfo
  {author} {\bibfnamefont {D.}~\bibnamefont {Vartanyan}}, \ and\ \bibinfo
  {author} {\bibfnamefont {H.}~\bibnamefont {Nagakura}},\ }\href {\doibase
  10.3847/2041-8213/ab191a} {\bibfield  {journal} {\bibinfo  {journal}
  {Astrophys. J. Lett.}\ }\textbf {\bibinfo {volume} {876}},\ \bibinfo {pages}
  {L9} (\bibinfo {year} {2019})},\ \Eprint {http://arxiv.org/abs/1812.07703}
  {arXiv:1812.07703 [astro-ph.HE]} \BibitemShut {NoStop}%
\bibitem [{\citenamefont {Powell}\ and\ \citenamefont {M\"uller}(2020)}]{m39}%
  \BibitemOpen
  \bibfield  {author} {\bibinfo {author} {\bibfnamefont {J.}~\bibnamefont
  {Powell}}\ and\ \bibinfo {author} {\bibfnamefont {B.}~\bibnamefont
  {M\"uller}},\ }\href {\doibase 10.1093/mnras/staa1048} {\bibfield  {journal}
  {\bibinfo  {journal} {Mon. Not. Roy. Astron. Soc.}\ }\textbf {\bibinfo
  {volume} {494}},\ \bibinfo {pages} {4665} (\bibinfo {year} {2020})},\ \Eprint
  {http://arxiv.org/abs/2002.10115} {arXiv:2002.10115 [astro-ph.HE]}
  \BibitemShut {NoStop}%
\bibitem [{\citenamefont {Obergaulinger}\ and\ \citenamefont
  {Aloy}(2020)}]{35OC}%
  \BibitemOpen
  \bibfield  {author} {\bibinfo {author} {\bibfnamefont {M.}~\bibnamefont
  {Obergaulinger}}\ and\ \bibinfo {author} {\bibfnamefont {M.~A.}\ \bibnamefont
  {Aloy}},\ }\href {\doibase 10.1093/mnras/staa096} {\bibfield  {journal}
  {\bibinfo  {journal} {Mon. Not. Roy. Astron. Soc.}\ }\textbf {\bibinfo
  {volume} {492}},\ \bibinfo {pages} {4613} (\bibinfo {year} {2020})},\ \Eprint
  {http://arxiv.org/abs/1909.01105} {arXiv:1909.01105 [astro-ph.HE]}
  \BibitemShut {NoStop}%
\bibitem [{\citenamefont {Nitz}\ \emph {et~al.}(2021)\citenamefont {Nitz},
  \citenamefont {Capano}, \citenamefont {Kumar}, \citenamefont {Wang},
  \citenamefont {Kastha}, \citenamefont {Sch\"afer}, \citenamefont
  {Dhurkunde},\ and\ \citenamefont {Cabero}}]{3-OGC}%
  \BibitemOpen
  \bibfield  {author} {\bibinfo {author} {\bibfnamefont {A.~H.}\ \bibnamefont
  {Nitz}}, \bibinfo {author} {\bibfnamefont {C.~D.}\ \bibnamefont {Capano}},
  \bibinfo {author} {\bibfnamefont {S.}~\bibnamefont {Kumar}}, \bibinfo
  {author} {\bibfnamefont {Y.-F.}\ \bibnamefont {Wang}}, \bibinfo {author}
  {\bibfnamefont {S.}~\bibnamefont {Kastha}}, \bibinfo {author} {\bibfnamefont
  {M.}~\bibnamefont {Sch\"afer}}, \bibinfo {author} {\bibfnamefont
  {R.}~\bibnamefont {Dhurkunde}}, \ and\ \bibinfo {author} {\bibfnamefont
  {M.}~\bibnamefont {Cabero}},\ }\href@noop {} {\  (\bibinfo {year} {2021})},\
  \Eprint {http://arxiv.org/abs/2105.09151} {arXiv:2105.09151 [astro-ph.HE]}
  \BibitemShut {NoStop}%
\bibitem [{\citenamefont {Abbott}\ \emph
  {et~al.}(2020{\natexlab{e}})\citenamefont {Abbott} \emph
  {et~al.}}]{GW190412}%
  \BibitemOpen
  \bibfield  {author} {\bibinfo {author} {\bibfnamefont {R.}~\bibnamefont
  {Abbott}} \emph {et~al.} (\bibinfo {collaboration} {LIGO Scientific,
  Virgo}),\ }\href {\doibase 10.1103/PhysRevD.102.043015} {\bibfield  {journal}
  {\bibinfo  {journal} {Phys. Rev. D}\ }\textbf {\bibinfo {volume} {102}},\
  \bibinfo {pages} {043015} (\bibinfo {year} {2020}{\natexlab{e}})},\ \Eprint
  {http://arxiv.org/abs/2004.08342} {arXiv:2004.08342 [astro-ph.HE]}
  \BibitemShut {NoStop}%
\bibitem [{\citenamefont {Mohamed}\ and\ \citenamefont
  {Jaïdane-Saïdane}(2009)}]{generalized}%
  \BibitemOpen
  \bibfield  {author} {\bibinfo {author} {\bibfnamefont {O.~M.~M.}\
  \bibnamefont {Mohamed}}\ and\ \bibinfo {author} {\bibfnamefont
  {M.}~\bibnamefont {Jaïdane-Saïdane}},\ }in\ \href@noop {} {\emph {\bibinfo
  {booktitle} {2009 17th European Signal Processing Conference}}}\ (\bibinfo
  {year} {2009})\ pp.\ \bibinfo {pages} {2273--2277}\BibitemShut {NoStop}%
\bibitem [{\citenamefont {Abbott}\ \emph {et~al.}(2018)\citenamefont {Abbott}
  \emph {et~al.}}]{o4_o5}%
  \BibitemOpen
  \bibfield  {author} {\bibinfo {author} {\bibfnamefont {B.~P.}\ \bibnamefont
  {Abbott}} \emph {et~al.} (\bibinfo {collaboration} {KAGRA, LIGO Scientific,
  Virgo, VIRGO}),\ }\href {\doibase 10.1007/s41114-020-00026-9} {\bibfield
  {journal} {\bibinfo  {journal} {Living Rev. Rel.}\ }\textbf {\bibinfo
  {volume} {21}},\ \bibinfo {pages} {3} (\bibinfo {year} {2018})},\ \Eprint
  {http://arxiv.org/abs/1304.0670} {arXiv:1304.0670 [gr-qc]} \BibitemShut
  {NoStop}%
\bibitem [{gwo()}]{gwopen}%
  \BibitemOpen
  \href@noop {} {}\bibinfo {howpublished}
  {\url{https://www.gw-openscience.org/}}\BibitemShut {NoStop}%
\end{thebibliography}%

\end{document}